\documentclass[tighten,times,twocolumn]{aastex631}
\usepackage{hyperref}
\usepackage{xspace}
\usepackage{graphicx}
\usepackage{booktabs}
\usepackage{comment}

\usepackage{amssymb,amsmath,epsfig}
\usepackage{subcaption}
\usepackage{floatrow}
\usepackage{mathrsfs}
\usepackage{xcolor}
\usepackage{mathtools}
\usepackage{relsize}
\usepackage{braket}
\usepackage[utf8]{inputenc}

\usepackage{makecell}

\newcommand{\tcm}{21$\,$cm\xspace}  

\newcommand{\aprx}{\mbox{$\ensuremath{\sim}$}}

\begin{document}

\title{Drone Beam Mapping of the TONE Radio Dish Array }

\author{Emily R. Kuhn}
\affiliation{Jet Propulsion Laboratory, California Institute of Technology, 4800 Oak Grove Drive,
Pasadena, CA 91109, USA}
\affiliation{Department of Physics, Yale University, New Haven CT 06520, USA}
\author{Will Tyndall}
\affiliation{Trottier Space Institute, McGill University, Montreal, QC, Canada}
\affiliation{Department of Physics, Yale University, New Haven CT 06520, USA}
\author{Benjamin R.~B. Saliwanchik}
\affiliation{Instrumentation Department, Brookhaven National Laboratory, Upton, NY 11973, USA}
\affiliation{Department of Physics, Yale University, New Haven CT 06520, USA}
\author{Anna Rose Polish}
\affiliation{Department of Physics, Yale University, New Haven CT 06520, USA}
\author{Maile Harris}
\affiliation{Department of Physics, Yale University, New Haven CT 06520, USA}
\author{Laura B. Newburgh}
\affiliation{Department of Physics, Yale University, New Haven CT 06520, USA}

\correspondingauthor{E. Kuhn and W. Tyndall}\\
\email{emily.r.kuhn@jpl.nasa.gov, william.tyndall@mcgill.ca}

\begin{abstract}



Drone-based beam measurements are a promising avenue to tackle the critical challenge of calibration for 21\,cm cosmology telescopes. In this paper, we introduce a new drone-based calibration system for 400-800\,MHz radio observatories, describing its instrumentation and first deployment. We discuss measurements of the TONE array, a CHIME/FRB outrigger pathfinder, and present results, including full 2D high spatial resolution beam maps in both co- and cross-polarization, as well as comparisons to simulations. The polarized beam maps cover a $70^{\circ} \times 70^{\circ}$ grid, capturing the first two sidelobes and measuring the TONE main beam and first sidelobe with $7-9\%$ statistical errors. We investigate polarization angle alignment with frequency, finding significant polarization leakage in the TONE antennas at frequencies above 600\,MHz, and a polarization axis rotation with frequency. We describe statistical and systematic errors, as well as measurements of radio frequency interference from the drone and equipment. Our drone system is the first to incorporate a broad-band switched calibration source in the drone payload, enabling background subtraction and direct measurements of the RFI emitted by the drone. The results presented are the first drone-based 2D measurements of cross-polar beam structure and of polarization alignment of an array. The high frequency and spatial resolution achieved with this system have revealed the rich structure of the beam of each antenna, and enabled comparisons between individual dishes and to electromagnetic simulations.


\end{abstract}

\section{Introduction}
\label{sec:introduction}

Current and future \tcm intensity mapping telescopes such as LOFAR~\citep{LOFARoverview}, MWA~\citep{MWAoverview}, CHIME~\citep{CHIMEoverview}, Meerkat~\citep{Meerkat}, CHORD~\citep{CHORD_Instrument}, HERA~\citep{HERAoverview}, HIRAX~\citep{2022JATIS...8a1019C,HIRAX_Instrument}, SKA~\citep{SKAoverview} and BINGO~\citep{Bingo2018} will measure the distribution of neutral hydrogen across a wide range of redshifts for a variety of science goals \citep{CV_Dark_Energy}. Sensitive measurements of the 21\,cm  power spectrum will measure the epoch of reionization and target large scale structure of galaxies, but significant challenges must be overcome to measure and remove bright synchrotron radio foregrounds from galactic and extragalactic sources. At high redshifts ($z>4$), experiments have benefited from identifying a subspace where the foregrounds are not mixed with the signal (`foreground avoidance') \citep{moraleshewittseparableforeground2004, tcmreionizationforegrounds, foregroundsubtractrequirementstcm, parsonsdelayspecfilterforeground, 
thyagarajanforegroundswidefieldspectra, dayenuforegroundfilter.500.5195E}, while at lower redshifts the cosmological signal and foreground have significant overlap in the equivalent subspace, necessitating alternative methods of foreground removal \citep{2020PASP..132f2001L}. Current analysis techniques rely on differentiating spectrally smooth foreground emission from the spectral structure of the \tcm signal which requires knowledge of the instrument beam to better than 1\% \citep{shaver1999, oh2003, liu2011, shaw2015coaxing}. 



The precision required for foreground removal is difficult to achieve with 21cm radio interferometers, many of which are stationary and cannot scan celestial sources to fully calibrate a beam pattern in 2D. While a variety of beam measurement techniques have been demonstrated \citep{HERA_beams,MWA_beams,CHIMEoverview, dallassolar, alex_SPIE,CHIME_holo}, they are limited in beam coverage and polarization information. The radio community has thus been developing techniques for beam mapping using unmanned aerial vehicles (UAV, hereafter drones) to address these challenges.  

Drone calibration is a particularly promising technique, and involves broadcasting a known calibration signal from a drone flying in a pre-determined 1D or 2D pattern over a telescope or array. This allows for control over the source location as well as its polarization, providing robust beam coverage for stationary telescopes and full polarization information for a broader class of telescopes. Over the past decade, drone-based beam calibration has become more viable due to rapid advances in drone systems, drone instrumentation, and the development of robust analysis software. Results from recent measurements include: 2D co-polarized beam maps of a 5\,m dish at Bleien Observatory with comparisons to solar transits \citep{Chang_drones} around 1420\,MHz; 2D co-polarized beam maps at 200\,MHz of a PAPER feed with promising comparisons to Orbcomm satellite beam measurements~\citep{Jacobs_drones,ECHO_update_drones}; 1D beam slices of a 3$\times$3 array of antennas at 408\,MHz with 0.5\,dB agreement between 3 measurements~\citep{Pupillo_drones}; 1D beam slices of a Tianlai dish with comparisons to the celestial source Cassiopeia A~\citep{Tianlai_drones}; and recent results which assessed drone measurements against outdoor antenna ranges, and found they were ultimately limited by reflections~\citep{Culotta-Lopez_drones}. Papers also demonstrate the breadth of applicability, including measurements in the Arctic~\citep{Herman_arcticdrone} and for oceanography receivers~\citep{Oceandrones}.

We developed and deployed a drone-carried payload to make beam measurements of \tcm telescopes between 400-800\,MHz. In this paper, we describe results from mapping the TONE radio array~\citep{TONE_instrument} at the Green Bank Observatory, which was built as a CHIME/FRB outrigger pathfinder~\citep{2024AJ....168...87L}. In Section~\ref{sec:instrument} we describe the drone, its payload, the TONE array, and the drone flights performed during beam measurements. In Section~\ref{sec:processing} we describe the analysis performed to produce beam maps of each of four dishes. In Section~\ref{sec:results} we present the results from this analysis, including co-polarized beam maps and statistical errors, beam widths, comparisons to simulations, cross-polarized maps and an investigation of polarization alignment. We will also discuss 
systematic errors, the largest of which comes from not accounting for the drone transmitter beam pattern. In Section~\ref{sec:rfi} we present anechoic chamber measurements of the radio frequency interference (RFI) from the drone, and compare it to in-flight data and requirements from current and future telescopes. Finally, we conclude in Section~\ref{sec:summary}. Although our best statistical errors are 7-9\% in the main beam and first sidelobe region, higher than the requirement of 1\%, our measurements represent the most complete far-field beam data set to-date from a drone-based calibration system.

\section{Instrumentation and Flights}
\label{sec:instrument}


The drone calibration system we have developed consists of a commercially produced drone platform (A DJI Matrice 600 Pro) augmented with a radiofrequency transmitting payload that produces a switched calibration signal between 400 and 800\,MHz in a single linear polarization. In this section, we describe the drone, the calibration payload, the TONE array, and the flights used in this analysis.

\subsection{Drone}
\label{sec:drone}


The drone platform we use is the Matrice 600 Pro produced by Shenzhen DJI Sciences and Technologies Ltd. (DJI) \footnote{\hyperlink{New flight plans can be uploaded while in flight, such that we can maximize the time we spend on a single set of batteries.}{DJI}}. The positional accuracy of the drone is improved significantly by a Real-Time-Kinematic (RTK) ground station that performs differential GPS measurements during flight. This system routinely achieves 20-30 minute flight times when flying at $\sim$200\,m altitudes with a 1.6\,kg payload attached. We utilize the mission planning and flight control software program UgCS\footnote{\hyperlink{https://www.sphengineering.com/flight-planning/ugcs}{UgCS: Flight Planning \& Control}} to generate flight-plans (described in Section~\ref{sec:flights}) and run on auto-pilot. New flight plans can be uploaded while in flight, such that we can maximize the time we spend on a single set of batteries.

\textbf{Communication:} Manually flying the DJI Matrice 600 Pro requires the use of a flight controller connected to a tablet and the RTK ground station. Control signals and telemetry data are transmitted at 2.45\,GHz between the drone and flight controller, and differential GPS measurements are transmitted between the drone and RTK ground station at 915\,MHz. To conduct a flight using the autopilot, a laptop running UgCS must be connected (via WiFi) to the tablet, which issues commands to the drone via the flight controller telemetry link. These additional digital electronics have the potential to introduce radiofrequency interference (RFI) into our measurements (see Section~\ref{sec:rfi}).

\textbf{Raw Data Processing:} The drone collects its own telemetry data in two forms, both of which can be accessed after flight, although we use only one of these data sets for this paper. Our primary data set is a high cadence ($\sim 5$\,Hz) data set that requires decryption. The encrypted data is decrypted and processed into CSV files using DatCon.\footnote{\href{https://datfile.net/}{datfile.net}} 
The data contains roughly 300 data fields, each written asynchronously at $\sim$5\,Hz, with the exact rate depending on the sensor. Sensors that read out at slower rates will result in repeated data points after processing by the DatCon software, and these points are removed later in analysis. The primary fields used are location in longitude, latitude, and altitude (height above mean sea level); orientation in three angles (pitch, roll, and yaw), velocity for each direction, as well as timestamps and status information. The timing columns provided are GPS time to the nearest second, and seconds since the drone was powered on, to the nearest millisecond. We combine these two columns to create a single interpolated time column that is synchronized to GPS time at a millisecond cadence.  


\textbf{GPS:}The drone system uses three different GPS modules. The first is a P-GPS that comes integrated into the drone. The second is the RTK GPS unit made by DJI to be compatible with the Matrice 600 Pro, and which we add for improved positional accuracy. However, most measurements presented in this paper do not use this RTK GPS position, as it was switched off for flights due to a faulty cable. The final GPS module is a UBlox RTK GPS, which is programmable and customizable. Although it can report location coordinates, we only use it to provide a timing lock output for the radio-frequency transmitter as described below.

\begin{table}[ht]
\centering
\begin{tabular}{|l|l|l|}
\hline
Drone sensor & Mean & 1$\sigma$ \\\hline
Return-to-Point (RTK on) & - & 2\,cm (specification: 1\,cm) \\
Return-to-Point (RTK off) & - & 1.5m (specification: 1\,m)\\
Height & 177\,m & $\pm$ 2.2\,m  \\
Yaw & - & $\pm 0.1^{\circ}$  \\
Pitch & $5^{\circ}$ & $\pm3^{\circ}$  \\
Roll & $2^{\circ}$ & $\pm1^{\circ}$  \\\hline
\end{tabular}\\
\caption{\label{tab:dracc} 
Cartesian location errors based on return-to-point measurements (first rows), and averages and standard deviations for height, roll, pitch and yaw during flight.
}
\end{table}

The location precision of the drone is summarized in Table~\ref{tab:dracc}. It was determined through a series of `return-to-point' measurements, in which the drone was placed on a common set of markers before and after each of the flights considered in this paper. The drone reports its position in absolute coordinates of latitude, longitude, and height. During data processing, a conversion to a local Cartesian reference frame\footnote{\href{https://pypi.org/project/PyGeodesy/}{PyGeodesy}} is performed using the latitude, longitude, and height of a known reference point (in this instance, the position of dish zero provided by surveying). The drone position is then converted to Cartesian and polar coordinate systems for each dish by subtracting their relative positions.  
For the RTK tests, we enforced a GPS location for the RTK ground station by hand, which was shown in lab tests to reduce jumps in position between flights. As a result, we cannot estimate the absolute position of the beam features within $\sim$1\,m for the flights presented in this paper.


The difference between the averaged start and averaged end positions was then computed ($r = \sqrt{\Delta x^{2} + \Delta y^{2}}$). We found the average difference for all flights is 2\,cm with the RTK on, and 1.5\,m with the RTK off. This shows clear improvement with the RTK, as expected. The average standard deviation of the height within a flight is 2.2\,m; height variations are corrected for during processing (as discussed in Section~\ref{sec:processing}). 

The average and standard deviation of the three angular coordinates (roll, pitch, yaw) are also given in Table~\ref{tab:dracc}. Roll and pitch refer to the tipping and tilting motion of the drone, with the pitch angle defined along the direction of travel of the drone; yaw angle is defined as the angle between the drone heading and North cardinal direction, and maps directly to the polarization angle of the transmitter (as the transmitter is affixed to the drone body). The pitch angle has an average value of $5^{\circ}$, which is expected because the drone body tips forward in order to fly at a constant velocity. As noted in Section~\ref{sec:source}, the broad beam of the transmission antenna ($\sim 50^{\circ}$) means that a $5^{\circ}$ tilt negligibly impacts our measurements, and thus we do not account for pitch in our analysis. The roll angle swings dramatically at turnarounds in the gridded flights, and so we excluded the turnarounds in the average and standard deviation calculations. This exclusion is done by considering only data where the velocity was greater than 4\,m/s or the roll angle was less than 5$^{\circ}$.  However, the standard deviation of the roll and pitch angles are still primarily driven by areas leading up to the turnarounds. The yaw angle is held constant during flights (the drone is programmed with a single heading), and is found to be stable to within $0.1^{\circ}$. Yaw deviations at this level will have a negligible impact on the transmitter power as measured at the telescope.

\begin{figure}[t]
    \centering
    \includegraphics[width=1.0\textwidth]{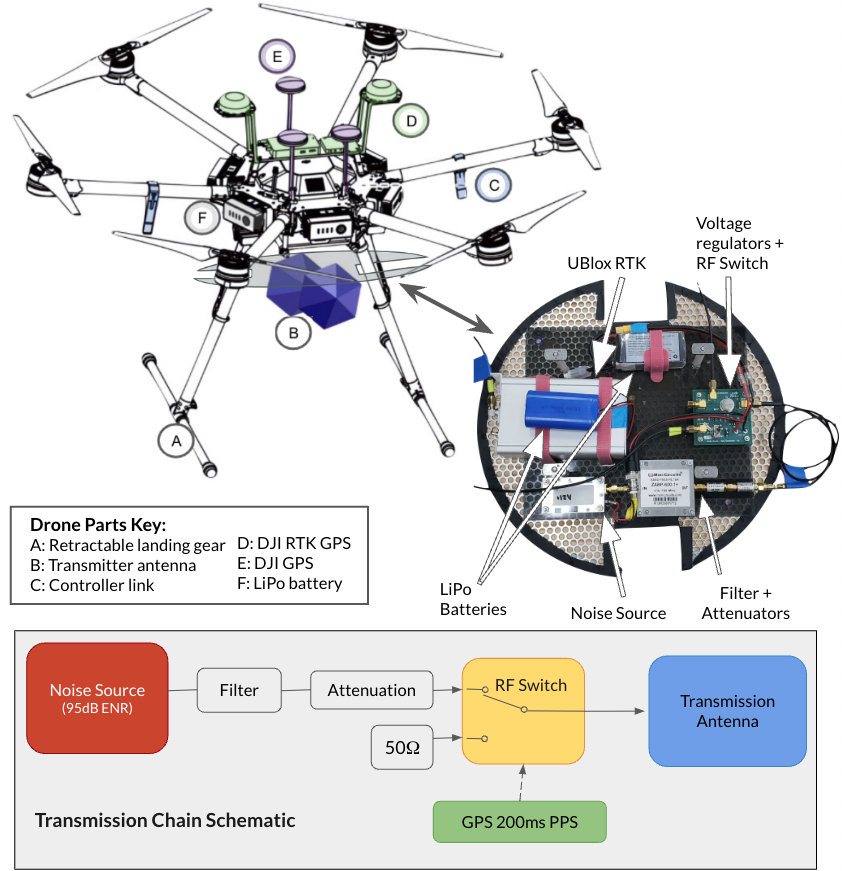}
    \caption[]{(Upper) A rendering of the Matrice 600 Pro, and a photograph of the payload with components indicated (described in more detail in Section~\ref{sec:drone}). The payload consists of the calibrator source, a GPS to provide a pulse synchronized to PPS, batteries for the source and components, and the transmission antenna.  (Lower) Schematic of transmission signal RF chain. }
    \label{fig:droneschem}
\end{figure}

\subsection{Drone Payload}
\label{sec:source}

The drone payload is shown in Figure~\ref{fig:droneschem} and consists of a 95\,dB ENR broadband calibration signal which is filtered to 400-800\,MHz band, attenuated down to 49--59\,dB ENR by analog components; a radio-frequency (RF) switch to toggle transmission between the noise source and 50\,$\Omega$ termination load; a UBlox RTK providing a synchronized signal for the switch; and a biconical transmitting antenna transmitting a single linear polarization aligned with the `yaw' angle of the drone. 
The source is switched to allow for subtraction of the background. The electronics are powered by two different 15V LiPo batteries and voltage regulators, which last for a full day of flights. The payload itself weighs 1.6 kg, and the noise source and transmission antenna are described in more detail below. 

\textbf{Noise Source:} The noise source is a commercial broad-band 
source, also used in \citep{Tianlai_drones}. It is filtered to the 400-800\,MHz band\footnote{\hyperlink{https://www.minicircuits.com/pdfs/ZABP-598-S+.pdf}{Minicircuits ZABP-598-S+ bandpass filter}}, and then sent to an RF switch\footnote{\hyperlink{https://www.minicircuits.com/pdfs/ZSDR-230+.pdf}{Minicircuits ZSDR-230+}}, all integrated into a custom PCB board. 
 
 The two inputs to the switching board are the filtered calibration source (`on' or `high' state) and a 50$\Omega$ terminator (`off' or `low' state). The switching is controlled with a pulse-per-second signal synchronized to GPS from the onboard uBlox ZED-F9P GPS system. For the calibration measurements of the TONE Array presented here, the GPS was configured to pulse at a 200\,ms cadence (200\,ms high, 200\,ms low), which corresponds to $\sim$4 samples in the high state and $\sim$4 samples in the low state. The switch isolation is $-40$\,dB on average across the band, providing an `off' signal that is $<1$\,K, smaller than the 100-400\,K antenna temperature of the TONE array~\citep{TONE_instrument}. The dominant signal in the `off' state appears consistent with RFI, as discussed in Section~\ref{sec:rfi}. In comparison, the signal in the `high' state is between 1000\,K (highest attenuation) and 10,000 \,K (lowest attenuation). The output of the switch is connected to the broadcast antenna, such that a pulsed signal is transmitted from the drone. 

In combination, the source and the RF switch have an overall bandpass slope of 1\,dB, which is within the target of 3\,dB for good dynamic range across the band for the TONE correlator. We measured the output spectrum from the calibration payload before and after most flights with a handheld spectrum analyzer \footnote{\hyperlink{https://www.rohde-schwarz.com/us/products/test-and-measurement/handheld/rs-fsh-handheld-spectrum-analyzer_63493-8180.html}{FS4 spectrum analyzer}}. We found that the source amplitude varied typically less than 3\% between six flights, although larger variations (6-8\%) were measured twice. The larger variations are on the same order as our statistical errors and thus source variability requires further attention in the future, although the contribution from the spectrum analyzer to this variation is not known. 



\textbf{Transmission Antenna: } The calibration signal from the noise source RF chain is broadcast with a biconical antenna from Aaronia\footnote{\hyperlink{https://aaronia.com/en/produkte/antennas/bicolog}{Aaronia BicoLOG 30100}}, selected because it is lightweight, broadband, linearly polarized, and has a broad beam. This style of antenna has been used by~\citep{Jacobs_drones} for similar applications. 
The biconical transmitting antenna is mounted on a ground plane which shields the drone RFI emission and reflections from the legs, and improves the forward gain of the antenna. This results in a smooth, broad beam across the full 400-800\,MHz frequency band. The beam pattern of the BicoLOG antenna mounted to a lightweight mock-up of the drone was measured
at the NCSU Nano Fabrication Facility anechoic chamber, and the beam of the BicoLOG mounted to the actual drone was measured on the Green Bank Antenna Test Range. 

Measurements of both co- and cross-polarization were made in two orthogonal cuts (one aligned with the polarization direction, one orthogonal) and show a broad beam, with full-width half-max (FWHM) values that range from 40$^{\circ}$-90$^{\circ}$ in the 400-600\,MHz frequency range used in this paper. At 450\,MHz, the frequency chosen for most figures, the FWHM is $\sim 50^{\circ}$. As noted in Sections~\ref{sec:mainbeam} and \ref{sec:polarized}, these beam measurements were not used to correct the drone measurements because they were limited to the two orthogonal cuts, and full 2D information was not available. In the future, such measurements of the transmitter should be performed and accounted for in the analysis. 





\subsection{TONE: CHIME/FRB Outrigger Pathfinder Telescope}
\label{sec:TONE_instrument}


\begin{figure}[t]
    \centering
    \includegraphics[width=1.0\textwidth]{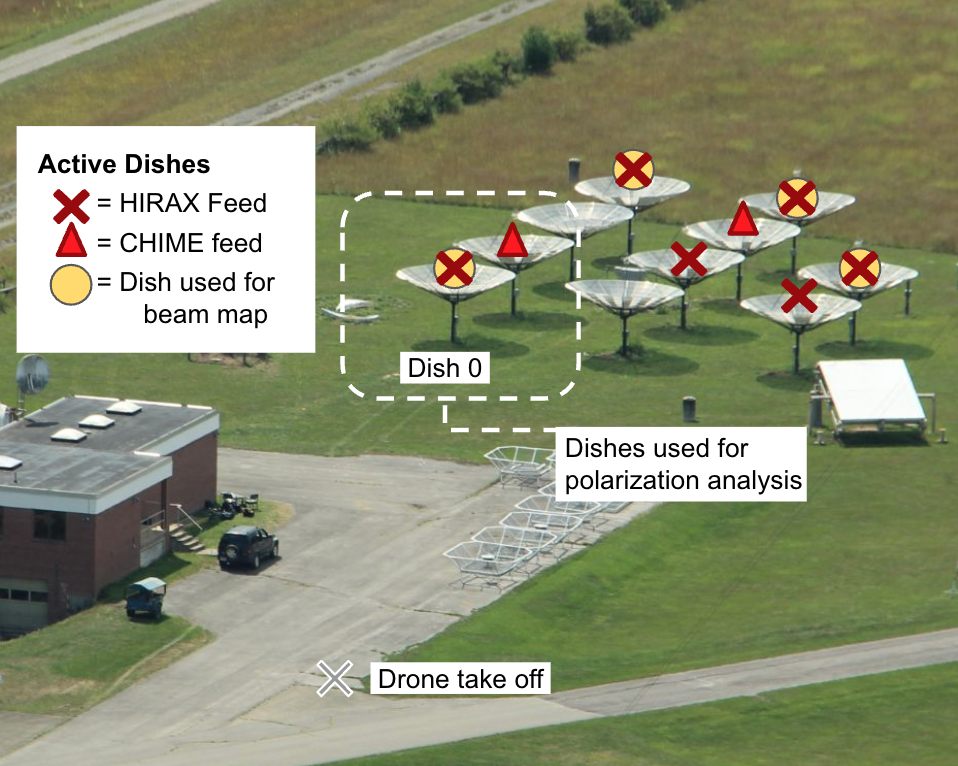}
    \caption[]{TONE at Green Bank Radio Observatory. The 8 active dishes are marked with X's, color-coded by whether the dish is instrumented with a HIRAX or CHIME feed. The four dishes used for drone beam mapping are marked with a yellow circle. The drone take off location and `Dish 0' location (which served as the coordinate system origin) are also marked. }
    \label{fig:tone_pic}
\end{figure}

The data presented in this paper come from flights performed over the TONE array~\citep{TONE_instrument}, the CHIME/FRB Outrigger Pathfinder in Green Bank, West Virginia, shown in Figure~\ref{fig:tone_pic}. The TONE array consists of eight 6\,m parabolic dishes, and operates in the 400-800\,MHz frequency band. The individual dishes are instrumented with either CHIME feeds~\citep{deng14} or HIRAX feeds~\citep{Kuhn_HIRAXantenna,saliwanchik21}, which provides a comparison between similar feed models. TONE was a local testbed for the HIRAX instrument because the dishes, analog instrumentation, and correlator are the same as the first HIRAX prototype array. 

The eight TONE dishes were manually pointed to zenith and rotated to align the two linear polarizations with cardinal directions (North-South and East-West) via hand-held compass, such that they would be oriented to GPS North within a few degrees. From fitting centroids to the drone data in this paper, we found the dishes could be mis-pointed up to 2$^{\circ}$ relative to the pointing for Dish 0 (see Figure~\ref{fig:tone_pic}). Because the RFI environment within the National Radio Quiet Zone (NRQZ) is so low, band-defining filters are not utilized in the TONE signal chain until after several stages of amplification. This was problematic for drone measurements because the DJI D-RTK base station and drone communicate through a frequency-hopping spread spectrum signal from 900-915\,MHz (see Table \ref{tab:rfi_table}) that aliases to 685-700\,MHz upon channelization. To address this, we inserted 400-800\,MHz band-pass filters into the front-end chains, prior to RF-over-fiber conversion. Despite these measures, when we examined auto-correlation data it became apparent that front-end amplifiers on Dish 2 were routinely driven into a non-linear oscillation that persisted until they were power-cycled. As a result, these channels are not included in the data presented here.

Ultimately we excluded data from 4 dishes from the analysis: Dish 2 was excluded because of the oscillations described above. Dish 3 (using a CHIME feed) and Dish 4 we excluded due to signal compression. Dish 6 had one of its inputs reserved for the clock signal and so was not connected. As a result, we restrict our analyses to the four remaining dishes with both North and East polarization data present. The exception is the rotation flight, where we use data from Dish 3 as a comparison because it was populated with a CHIME feed (see Section~\ref{sec:rotate360}). 

The analog signals from all inputs are digitized, channelized, and correlated using a 16-channel ICEboard FPGA system as described in~\citep{ICEboard}. The data is channelized into 1024 frequency bins between 400-800\,MHz in the second Nyquist zone, with an integration rate selected to provide the fastest possible data cadence, resulting in output frames that are $41.94304$\,ms ($2^{14}\times2.56\,\mu\textrm{s}$) in duration. For each correlator frame we obtain the full $N^2$ visibility matrix from the ICEboard for N=16 inputs (120 correlation products). Only the auto-correlation products are shown in this paper, and thus all beam measurements presented are in units of power.

The ICEboard applies a per-frequency gain after channelization to reduce quantization noise. Typically these gains are set such that the quiet sky is in the middle of the bit range of the correlator. Instead, we fixed the gain such that the signal level from the quiet sky appeared at a low digital power of $\sim1000$ (out of $2^{16}=65536$) across the full frequency band, corresponding to $\sim1.5\%$ of the correlator bit-depth of each 42\,ms integration frame. This amplitude was sufficient for performing background subtraction, and left a majority of the bit-depth ($\sim98.5\%$) available for measuring the bright drone calibration signal. Thus we could broadcast the signal at a higher amplitude before digital saturation occurred than if we had used the settings optimized for measuring the quiet sky.

Together, these correlator configuration choices for data integration interval and gain optimization permitted high signal to noise at 42\,ms cadence, allowing us to measure the far-field beam pattern with the drone within a single flight.

These drone flights were conducted during an extended maintenance window for the TONE array. During both flight campaigns, a transformer that provided power to the buildings that housed the TONE correlator, precision GPS clock, and network infrastructure was not operational. Backup generators supplied to power these devices during daylight hours, but their performance was intermittent and at times unreliable. Due to the power interruptions, the GPS clock was unable to reach sub-second accuracy (relative to UTC) which was discovered in post-processing.

\subsection{Drone Flights} 
\label{sec:flights}

The drone measurements produce 2D telescope beam maps in two polarizations that include both main beam features and sidelobes. To build these maps, we flew two different flight types at an altitude of 180\,m, corresponding to the far field distance at $\sim$ 750\,MHz: (1) Flights transmitting higher power ($\sim$60dB ENR), which will capture far side-lobe features but saturate the main beam (the result will look like a flat-topped Gaussian when measuring a parabolic dish); (2) Flights transmitting lower power ($\sim$50dB ENR), which give resolution in the main beam but do not illuminate far side-lobes above the background. Amplitude levels between the two styles of flight can be re-scaled and matched to build a complete map. To ensure beam coverage and for ease of autopilot programming, we fly the drone in a planar grid. Grid spacing is optimized to fit as many passes as slowly as possible within the defined grid area within the charge of a single set of batteries ($\sim$ 30\,minute flight time). 

We began by choosing the attenuation levels for the main beam and sidelobe flights by manually centering the drone by-eye while it hovered over the dish, and then sending the drone to the far-field height, rotating 360$^{\circ}$, and inspecting the autocorrelation time stream data for flat-top features that might indicate compression. We found attenuation levels of -46\,dB for main beam flights and -36\,dB for sidelobe beam flights were appropriate.

\begin{table}[t]
\centering
\begin{tabular}{|l|l|l|l|l|l|l|}
\hline
Flight & Head & grid & width & Att. & RTK & $\Delta t$ \\
 & & dir. & [m] & [dB] & [Y/N] & [s] \\\hline
618 & N & NS & 100 & -46 & No & -1.5 \\
619 & N & EW & 100 & -46 & No & -1.4 \\
625 & N & NS & 200 & -46 & No & -1.26 \\
647 & N & EW & 200 & -40 & No & -3.15 \\
646 & N & NS & 200 & -40 & No & -4.57 \\
533 & N & EW & 200 & -36 & Yes & 0.02 \\
623 & S & NS & 100 & -46 & No & -3.26 \\
620 & E & NS & 100 & -46 & No & -3.51 \\
649 & E & EW & 200 & -40 & No & -3.37 \\
648 & E & NS & 200 & -40 & No & -3.45 \\
535 & E & EW & 200 & -36 & Yes & -1.35 \\
608 & rot. & rot. & - & -46 & No & $\sim3.7$ \\ \hline
\end{tabular}\\
\caption{\label{tab:flights} Flight Summary. Indicated are the flight numbers, the heading which also corresponds to polarization orientation, the direction of flight and width of the grid (100\,m is a main beam flight, 200\,m is a sidelobe flight), the attenuation on the payload, whether the RTK was operational, and the time stamp offset between the drone and the data as found during processing. The last flight indicated was a series of 360$^{\circ}$ rotations for polarization analysis. One flight (623) had a South-oriented angle to check for systematics. }
\end{table}

Data presented in this paper was obtained over two week-long drone flight campaigns in August and October 2021. We flew 48 flights, of these we analyzed data from 12 flights. 14 flights were not used in the analysis presented here because they were at other polarization angles, 13 flights were excluded due to data dropouts before changing the data acquisition drives from hard-drives to solid state drives, four flights were for RFI testing, two flights only had a few telescope inputs working, two flights had odd behavior (likely the RF source losing battery power), and one 'main beam' small flight showed significant compression and could not be used. Of the 12 flights presented here, 10 were gridded flights with the polarization of the drone oriented either North or East, one had the drone polarization oriented South, and one was a 360$^{\circ}$ rotation.  

\begin{figure*}[th!]
\centering
\includegraphics[width=1.0\textwidth]{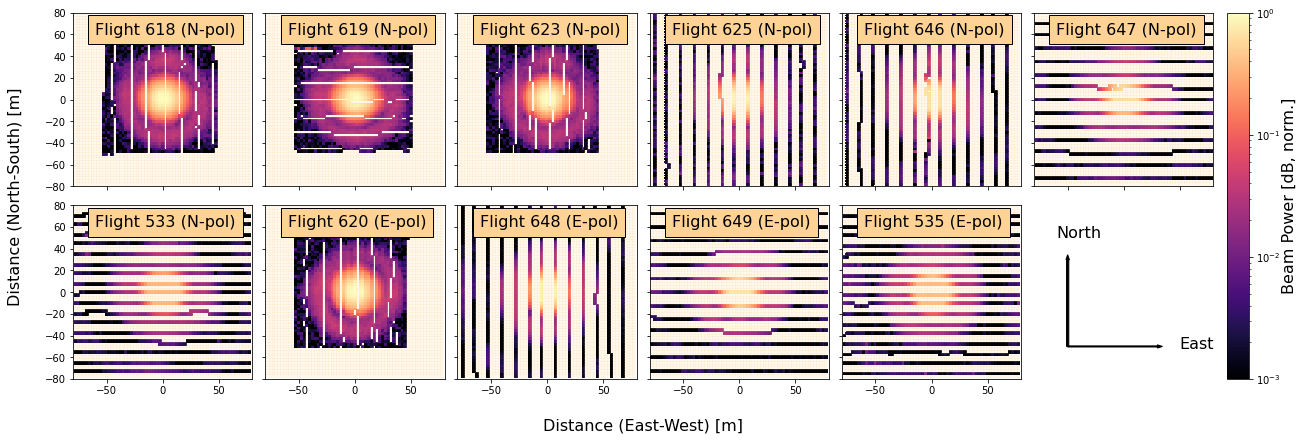}
\caption[]{Each panel shows the beam measured from each of the flights used in this analysis for a single dish and frequency after the per-flight processing performed in Section~\ref{sec:processing}.  The cardinal direction `North' is vertical; some flights were oriented North-South and others East-West, and color indicates beam amplitude. The polarization during the flight was either aligned North-South (`N-pol') or East-West (`E-pol'). The smaller grids target a measurement of the main beam, the larger grids target a measurement of sidelobes. Additional detail on the flights is given in Table~\ref{tab:flights}. 
}
\label{fig:flight_summary}
\end{figure*}

An overview of these flights is presented in Table~\ref{tab:flights}, and the data from each flight for a single frequency and dish is shown in Figure~\ref{fig:flight_summary}. In what follows, we choose a convention of a heading of `North', where the polarization axis should be aligned North-South, to be 0$^{\circ}$. We chose the size of the grid separately for sidelobe flights and main beam flights. The brighter sidelobe flights were 200\,m$\times$200\,m, allowing us to measure sidelobe features out to $\sim$40$^{\circ}$. The dimmer main beam flights used a grid size of 100\,m$\times$100\,m (out to $\sim$15$^{\circ}$) to measure the main beam and first sidelobe. Grid direction refers to the direction of the passes over the telescopes. For example, for a NS 100\,m width grid the drone would fly a pass 100\,m across the telescope from north to south, shift \aprx2\,m east, and fly 100\,m south to north, and repeat to fill a 100\,m $\times$ 100\,m field. Since all data taken is more finely sampled along the direction of the grid passes, to ensure adequate and uniform spacial sampling NS and EW grids were flown for each beam map. Thus, a single measurement set requires 8 gridded flights: two polarizations, two grid directions, and two grid widths, to cover both the main beam and sidelobes. 

\section{Data Processing }
\label{sec:processing}



In this section we describe the pipeline developed to process the raw data from the drone and ICEBoard, identify the flashing drone signal, solve for and apply timing corrections, grid the data into maps, normalize the sidelobe and main beam flights, and co-add flights together for each polarization.  

\subsection{Background Subtraction}

The broad-band white-noise calibration signal emitted by the drone is pulsed with a period of $400$\,ms with a 50\% duty cycle (as described in Section \ref{sec:instrument}). Every period results in $\sim4$ correlator frames where the pulsed source is on, $\sim4$ correlator frames where the source is off, and $\sim2$ frames that contain the rising and falling edge of the pulsed signal. 

During the data reduction process, we identify the pulsing behavior for each correlator frame through an iterative fitting process. We construct a square wave $S(t)$, which has an on value of 1.0 and an off value of 0.0, with the same period $T$ and duty cycle (50\%) as the pulsing calibration signal:

\begin{equation}
    S(t)=\frac{1}{2}\textrm{sgn}\left[\textrm{sin}\left(\frac{2\pi(t-t_0)}{T}\right)\right]+\frac{1}{2}
\end{equation}

The value $t_{0}$ is the time offset of the square pulse in the data which we solve for by iterating through one pulse period of offset in 1\,ms steps. For each time step, we compute the Pearson-R (PR) correlation between the square wave and the auto-correlation data for one frequency channel. The time offset $t_0$ that maximizes this PR correlation parameterizes the best-fit square wave.

\begin{figure}[b]
    \centering
    \includegraphics[width=1.0\textwidth]{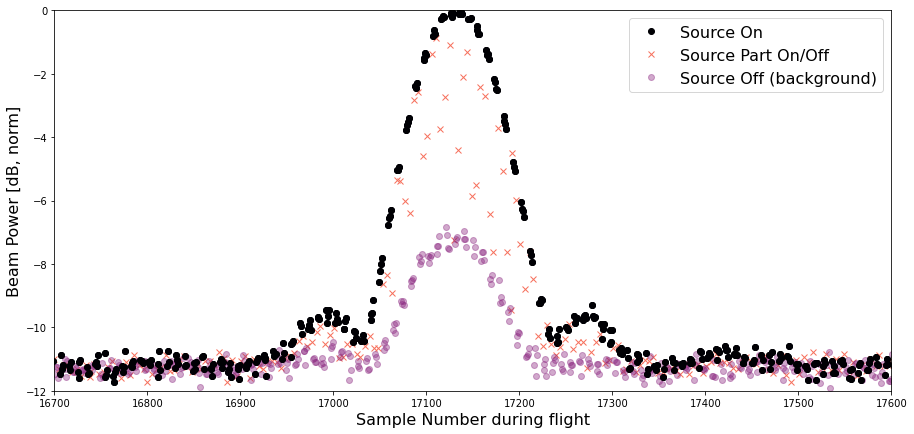}
    \caption[]{A single transit of the drone through near the center of the beam for Dish 0, North-polarized input from  flight 625 at frequency 448\,MHz. Indicated by color are points in the flight where the background fitting found the signal was on (black), off (purple), or partially on and partially off (peach). Only `On' data is used in the following analysis after background subtraction. }
    \label{fig:bg_slice}
\end{figure}

The timestamps of each correlator frame are then interpolated to the best-fit square wave at 1\,ms resolution, much higher than their native 42\,ms cadence, to identify the fraction of calibration signal for each correlator frame. Using this interpolated data set, we find 42\,ms data frames with only `on' data ($S(t)=1$), only `off' data ($S(t)=0$), and frames that contain a mixture (some 1\,ms samples on, some off within the 42\,ms frame). Correlator frames that contain the rising or falling edge of a calibration pulse are omitted from analysis to avoid biasing the background or drone signal arrays. Figure \ref{fig:bg_slice} demonstrates the successful sorting of the correlator frames using this method. In that figure, the source-off data clearly has a peak at the Gaussian peak of the source on signal, we believe this is primarily due to RFI from the drone (see Section~\ref{sec:rfi}).

The total power, $P_{\textrm{meas}}(t_{i},\nu)$, measured at time $t_{i}$ in frequency bin $\nu$ is the sum of the drone calibration signal $P_{\textrm{drone}}(t_{i},\nu)$ and several background signals, which include the drone RFI noted above, the sky, and the correlator noise floor:

\begin{equation}
P_{\textrm{meas}}(t_{i},\nu)=P_{\textrm{drone}}(t_{i},\nu)+P_{\textrm{bkg}}(t_{i},\nu)
\end{equation}

To isolate the white-noise calibration signal $P_{\textrm{drone}}(t_{i},\nu)$ in each auto-correlation measurement $P_{\textrm{meas}}(t_{i},\nu)$, we subtract an estimate of the background $\hat{P}_{\textrm{bkg}}(t_{i},\nu)$. When the calibration signal is off ($P_{\textrm{drone}}=0$), the measured power is the background itself. When the calibration signal is on ($P_{\textrm{drone}} \neq 0$), the background at that timestep is estimated from a rolling average of twenty background data points, ten on each side of $t_{i}$. We found ten points on each side approximated the background without being too impacted by transient spikes and retaining relevant changes such as extra background power near the center of the beam.

This operation is performed for all correlator frames ($t_{i} \sim \mathcal{O}(10000)$) across the full frequency axis (1024 elements) for all auto-correlations (16 channels) in the $N^2$ visibility matrix acquired during each drone flight. The result is a background signal for each frame, frequency, and input, taken as the measured background when the calibration source is off, and the rolling average estimate when the source is on. The rolling average estimate can now be subtracted from the source-on data: $P_{\textrm{drone}}(t_{i},\nu) = P_{\textrm{meas}}(t_{i},\nu) - \hat{P}_{\textrm{bkg}}(t_{i},\nu)$

We found that the time-dependent background computation method used in this paper performs poorly when fast ($\mathcal{O}(1)$ frame) RFI features are present because the computed background will be biased high for all frames that contain this RFI contaminated background measurement. However, it successfully mitigates background level changes caused by slowly varying RFI ($\mathcal{O}(10)$ frames), the increased power caused by the drone entering the antenna field of view ($\mathcal{O}(100)$ frames), and gain drifts over longer timescales ($\mathcal{O}(1000)$ frames).

\subsection{Timing Correction}

A timing offset between the correlator and drone timestamps is apparent in uncorrected 2D beammap plots. As previously mentioned (see Section \ref{sec:instrument}) the GPS-disciplined clock system providing UTC timestamps to the TONE data acquisition computer was not consistently powered, resulting in inaccurate timestamps being stored with the correlator frames. We reconstruct the correlator time axis from the FPGA frame count (which is precisely controlled by an internal oscillator) and the first timestamp sent by the compromised external clock. The reconstructed time axis precisely matches the duration of each correlator frame, but contains an offset from UTC (and thus the drone) which must be corrected.

For each drone flight, we calculate (and rectify) the timing offset present between the correlator and drone data. We apply an offset $\Delta t$ to the correlator timestamps, reinterpolate the drone data to the shifted correlator timestamps, and perform a 2D Gaussian fit on the reinterpolated data. The Pearson-R correlation between the best-fit 2D Gaussian and the reinterpolated data is calculated for each iteration. We first iterate across a 20\,s range $-10\,\textrm{s}\le\Delta t\le+10\,\textrm{s}$ with coarse resolution ($\delta t=0.2$\,s) to find the offset $\Delta t_C$ that maximizes the PR value. We then iterate across a 1\,s range $\Delta t_C -0.5\,\textrm{s} \le\Delta t\le\Delta t_C+0.5\,\textrm{s}$ with fine resolution ($\delta t=0.01$\,s) to find $\Delta t_f$---the offset we apply to synchronize our data products for analysis. All offsets are given in Table~\ref{tab:flights}; we find timing offsets up to 4.5\,s with typical values between 1.5-3.5\,s. After applying the correction, the correlator and drone data are synchronized and the drone positions have been interpolated to the 42\,ms data cadence of TONE. This yields a co-sampled beam map (for all frequencies and inputs) for each flight.

\begin{figure*}[t]
    \centering
    \includegraphics[width=1.0\textwidth]{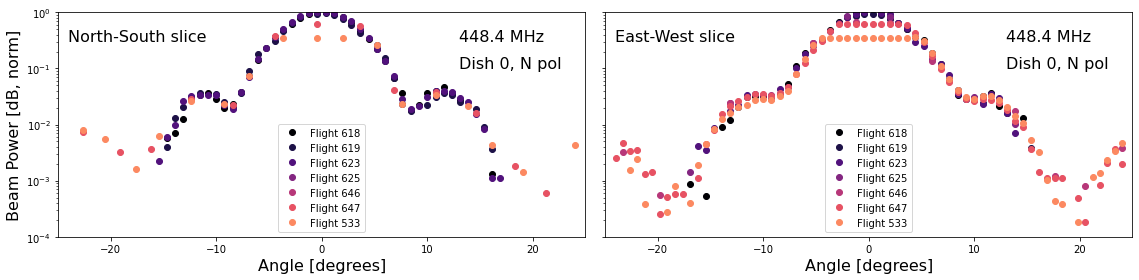}
    \caption[]{Beam power vs angle in a single 1D cut through the beam at the origin for frequency 448 \,MHz and a North-polarized input. (Left) shows a 1D cut through the North-South direction and (Right) shows a 1D cut through the East-West direction. Each flight is shown as a different color after the normalization correction procedure described in Section~\ref{sec:ampcorr}. As described, the amplitude correction procedure matches the average signal power in the first sidelobe region. As a result, the main beam flights have good signal in the main beam and the sidelobes, while the brighter sidelobe flights will look flat within the main beam where the signal was saturated. }
    \label{fig:normalization_slices}
\end{figure*}

\subsection{Gaussian fitting}
\label{sec:gaussfit}

For each flight, the beam map is fit with a 2D Gaussian in Cartesian coordinates (x,y) for each input and frequency:
\begin{equation}
G(x,y) = A e^{ - \frac{(x - x_{0})^{2}}{2 \sigma_{x}^{2}} - \frac{(y - y_{0})^{2}}{2 \sigma_{y}^{2}}}  + C
\end{equation}

Where $A$ is the amplitude, $x_{0}$ and $y_{0}$ are the beam centroids, $\sigma_{x}$ and $\sigma_{y}$ are the beamwidths (such that the full-width-half-max, FWHM, is $ 2\sqrt{2\mathrm{ln}2} \sigma$), and C is an offset. The amplitude is used to normalize the flights, and the centroids are used to center each flight per input, while the widths and offsets are not used in further processing. When we included an additional fit parameter representing the preferential angle of the semi-major and semi-minor axes of the beam, we found shifts in the FWHM parameter of up to $\pm0.2^{\circ}$ compared to fits without this additional parameter. As expected, larger preferential angles corresponded to larger FWHM shifts. However, the FWHMs jumped unphysically with frequency when we included this parameter. As a result, we do not present the fits with the additional parameter, but include $\pm0.2^{\circ}$ in the FWHM error bars, which is the dominant source of error in the FWHM.

\subsection{Gridding}
\label{sec:gridding}

Next, we bin the background subtracted data into a common grid across flights. This gridding can be performed in Cartesian coordinates with grids oriented along North-South and East-West and pixels that have separate user-defined widths and spacings in the North-South and East-West directions. We can also initialize the grid in polar coordinates $\theta$ and $\phi$; $\theta$ is counter-clockwise from East and $\phi$ defined as the angle from the zenith of the dish ($\phi=0^{\circ}$) to the horizon ($\phi=90^{\circ}$). In polar coordinates, the pixel widths are defined as constant $\theta$ and $\phi$, and thus the pixels are not a uniform shape in the projections in this paper. For a given flight and input, the fit centroid value in $(x,y)$ is subtracted from the flight drone data points as a function of frequency; this moves the center of the Gaussian beam to the center of the grid. 

Each time-domain value has been scaled with a radial distance compensation coefficient to account for the $1/r^{2}$ drop off in brightness received by the drone. The factor applied is $(r(t)/r_{0})^{2}$ where $r_{0}$ is the minimum distance between the drone and the surveyed location of the dish achieved during a given flight, while $r(t)$ is the time-dependent distance, each computed as $r = \sqrt{x^{2} + y^{2} + h^{2}}$. This effect is around 4\% at the first sidelobe, and can be as large as 50\% at 35$^{\circ}$, the largest angles presented here. This distance compensation correction is applied per-time sample, which also corrects for the altitude variations $\mathcal{O}[1]$\,m that can occur within a single flight. Optionally, a normalization can be applied to each flight at this stage. 

Gridding can be performed for flights individually, or a coadded map can be created from multiple flights that combines all telescope data before grid sampling. Telescope data is accumulated within each pixel, and the number of data points, average, and standard deviation is computed per pixel. This polar pixelization doesn't preserve constant area, and so pixels near the center are smaller. Consequently, central pixels contain fewer samples (often fewer than 5 counts per pixel) and are thus excluded from the calculation of standard deviation. This pixelization could be revisited for future processing efforts, since the coverage is fairly uniform when Cartesian gridding is used.

\subsection{Normalization between Flights}
\label{sec:ampcorr}

Ideally the Gaussian fits would provide the normalization for each flight, and could thus be used without modification for co-adding all gridded flights together. However, while we found that the flights with the most attenuation (46dB, main beam flights) had peaks with no evidence of compression, we also found that the less attenuated flights (40dB and below, sidelobe flights) had flat tops in the main beam associated with saturation of the correlator ADCs. For these saturated flights, the Gaussian fits would over-predict the height of the main beam, causing an amplitude discrepancy everywhere in the beam between flights at different attenuations. To correct these Gaussian amplitudes, we chose to match the amplitudes between flights using the first sidelobe region. We chose this region because it had sufficient signal to noise in both main beam and sidelobe flights, and was not biased by compression in the sidelobe flights. 

To perform this matching, we first chose a 46dB main beam `reference' flight for each polarization (flight 618 for the North polarizations and flight 620 for the East polarizations) and normalized the beam maps by the best-fit Gaussian amplitude, such that the peak value of those flights is defined to be 1.0 for each input and frequency. We gridded each data set onto a Cartesian grid of 2.5\,m$\times$2.5\,m, the result is a centered beam map on a common grid per frequency and input, normalized by its best-fit Gaussian amplitude. We scaled each non-reference flight by a range of values from 0.1 to 3.0 in 0.01 steps and subtracted this scaled map from the reference flight map, resulting in a differenced map between the reference flight and scaled flight for each input and frequency. The absolute value of the median of pixel values in a Cartesian annulus between 20-40m (for North-polarized inputs) or 18-35\,m (for East-polarized inputs) was computed for each value of the scaling. If the scaling were perfect and the beams matched exactly, the median value should be zero. The scaling that produced the minimum median value was selected as the correction to the Gaussian amplitude.


We found that the correction for other 46\,dB unsaturated main beam flights was small ($\sim$0.95-0.98). For the brighter, compressed flights the optimal correction values were less than 1.0 (ranging from 0.2 - 1.0, with medians between 0.45 - 0.8), indicating that the best-fit Gaussian amplitudes were over-estimating the amplitude. The resulting corrections to the Gaussian fit values did not always yield good overlap between main beam and sidelobe flights for all frequencies and all dishes. To maximize the number of flights we could include in the resulting data set, we chose to impose a frequency-cut to remove any frequencies that, for any dish and flight combination: (a) hit a minimal (0.1) or maximal (3.0) value during the median minimalization, indicating an optimal least-squares solution was not found during fitting, or (b) had a median differenced value greater than 6E-5 (a value chosen by visual inspection) indicating that the map differences in the annular region are large and hence the sidelobes may not match well. Only 141 of 512 frequencies passed this cut for all flights. The vast majority of the cut frequencies were removed by a single flight: 648. When we excluded flight 648, we found 322 frequencies passed the cuts instead. We identified that the digital gains of the TONE correlator were set too low during flight 648, such that for a wide range of frequencies the auto-correlation visibilities only exceeded zero when the drone was directly over a given receiver. Moreover, the background measurements at these frequencies were effectively zero throughout flight 648, and were thus improperly characterized when the drone was flying through the sidelobe regions of the receiver beams.

Figure~\ref{fig:normalization_slices} shows a horizontal and vertical slice through the center of the beam for each flight after applying the corrections to the Gaussian normalization. The sidelobe amplitudes match well for all flights, as would be expected from our normalization scheme. Within the main beam, low amplitude flights trace the Gaussian main lobe, while high amplitude flights (intended to map the sidelobes) appear as flat lines---evidence of saturation. The improved signal to noise of the high-amplitude flights is evident from the presence of a second sidelobe at $\sim \theta=25^{\circ}$.

Finally, we generated a plot similar to Figure~\ref{fig:normalization_slices} for each of the 141 frequencies that passed our initial criteria, and through visual inspection we identified 16 additional frequency bins that were unsatisfactory. Our final frequency axis contains 125 frequencies that are used in the following analysis.

\begin{figure}[t]
    \centering
    \includegraphics[width=1.0\textwidth]{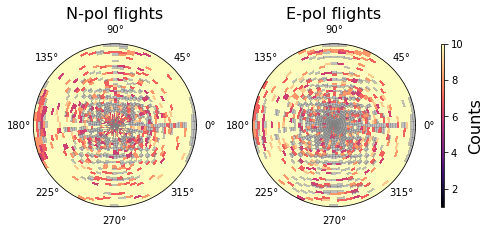}
    \caption[]{Counts per pixel (`hit map') after gridding, shown in polar coordinates at 448\,MHz. (Left) shows the East-polarized input for Dish 0 and (Right) shows the North-polarized input for Dish 0. The hit map is input and frequency dependent. In polar coordinates counts range from 0-30, typical values are 15. There are fewer East-polarized flights, and so those hit maps are sparser than the North-polarized flight hit maps. We mask pixels with fewer than 5 counts to indicate where the standard deviation cannot be calculated.}
    \label{fig:flightcounts}
\end{figure} 

\subsection{Coadding}

Our final processing step is to co-add data from each flight, grouped by transmitted polarization, using the gridding (see Section~\ref{sec:gridding}) and normalization (see Section~\ref{sec:ampcorr}) procedures defined previously. The result is a unique beam map for each dish, polarization, and frequency. We perform the coadding process twice to produce beam maps in both Cartesian and polar coordinates. The Cartesian grids are square, ranging from -80\,m - 80\,m with 2\,m spacing in each dimension. The polar grids span $360^{\circ}$ in $\theta$ in steps of $6^{\circ}$ and $36^{\circ}$ in $\phi$ in steps of $1^{\circ}$. To exclude compressed or low signal-to-noise regions we mask the high source amplitude flights within 18\,m ($6^{\circ}$ in polar coordinates) of the origin, and the low source amplitude flights outside of 40\,m ($12^{\circ}$ in polar coordinates) from the origin. Within each grid pixel, we combine data from all flights prior to computing the average and standard deviation per pixel. The resulting hit map for all flights in polar coordinates is shown in Figure~\ref{fig:flightcounts}. In Cartesian grids, the hit map is uniformally spaced because each pixel is the same size. Conversely, the polar grid cells (see Figure~\ref{fig:flightcounts}) do not have equal areas, so the central pixels are sparsely populated while the edges are densely sampled.

\section{Beam Patterns from Drone Measurements}
\label{sec:results}

In this section, we present the beam maps and describe the main beam, sidelobe, polarization, and systematics measurements from the 12 flights described in Section~\ref{sec:flights}, for the four TONE dishes (8 dual-polarized inputs) and 125 frequencies remaining after data selection. In maps or other plots in which a frequency was selected, we have chosen to show a representative frequency of 448.4\,MHz. In plots that emphasize frequency-dependence, we typically present data from dish 0, which resides at the nominal center of each drone flight. Because our presented measurements are restricted to autocorrelations, all amplitudes are in units of power.  Because the transmitting antenna is fixed to the drone body and only planar flights were performed, these measurements are equivalent to the Ludwig-I convention of polarization measurement, compared to the conventional Ludwig-III definition which is defined on a sphere \citep{ludwig73}. We thus over-predict the sidelobe levels by 4\% in the first sidelobe and 20\% in the second sidelobe, relative to the Ludwig-III definition. 


\subsection{Beam Maps}
\label{sec:mainbeam}

\begin{figure*}[h!]
    \centering
    \includegraphics[width=0.93\textwidth]{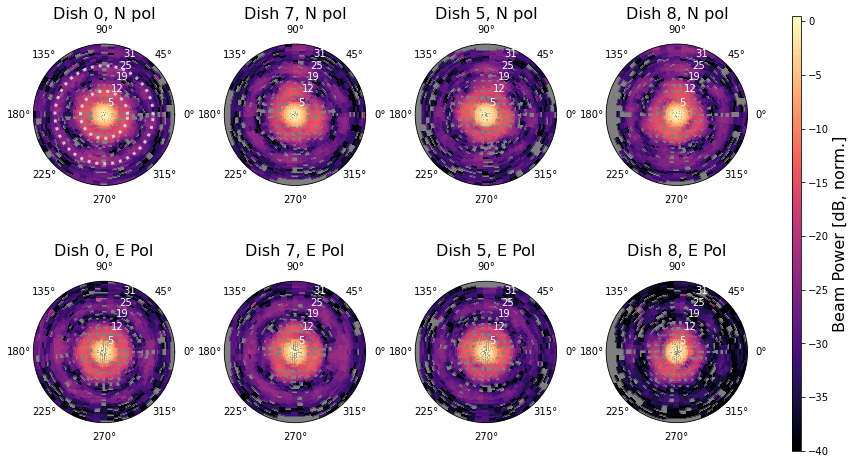}
    \caption{(Top row) shows 2D beam measurements in polar coordinates for North-polarized inputs for all four dishes considered in this paper at frequency 448\,MHz after processing (see Section~\ref{sec:processing}). (Bottom row) shows 2D beam measurements for East-polarized inputs.  Circles overlaid at $12^{\circ}$ and $25^{\circ}$ indicate the approximate locations of the first and second sidelobes. The data is normalized, and shown in logarithmic scale. }
    \label{fig:theta_phi_beam}
\end{figure*}

\begin{figure*}[h!]
    \centering
    \includegraphics[width=0.93\textwidth]{ 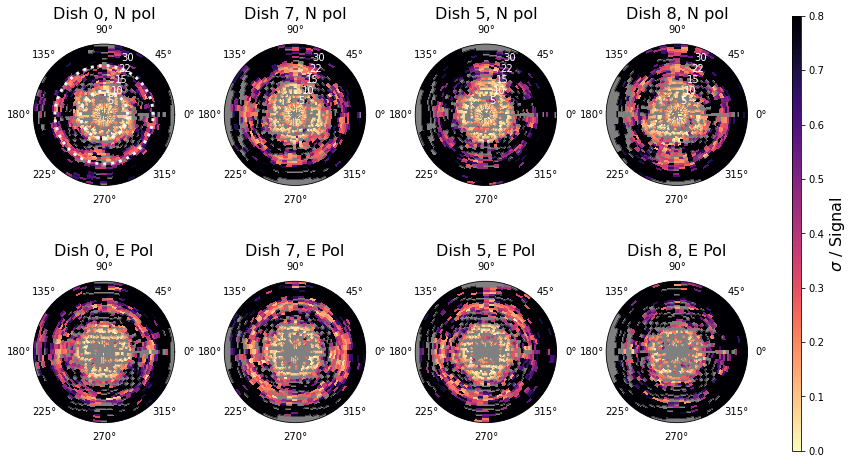}
    \caption{(Top row) Standard deviation per pixel, divided by the value of the pixel, for all North-polarized inputs, shown in polar coordinates. (Bottom row) Standard deviation divided by the value of the pixel, for the East-polarized inputs. Pixels with fewer than five counts have been excluded. The color scale has been inverted compared to Figure~\ref{fig:theta_phi_beam} to better highlight the regions with lower noise-to-signal (orange and yellow regions). We find $10\%$ variations in the main beam and first sidelobe regions, and $25\%$ variations in the second sidelobe region.  
    }
    \label{fig:errors}
\end{figure*}

\begin{figure*}[ht!]
    \centering
    \includegraphics[width=1.0\textwidth]{ 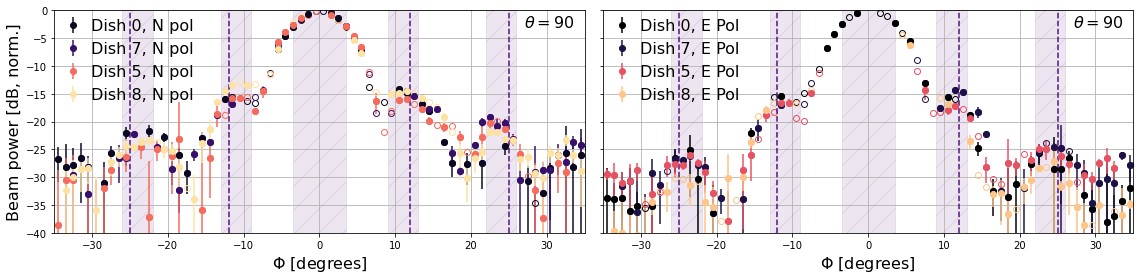}
    \caption[]{(Left) Slices through the beam at the origin, all dishes, at 448\,MHz, for the North-polarized inputs. (Right) Slices through the beam for the East-polarized inputs. Beam power data are shown as filled circles with $1\sigma$ errors. Open circles indicate data with 5 or fewer data points, and thus we do not report statistical errors. Shaded regions for the the main beam ($\phi < \pm3.5^{\circ}$), first sidelobes ($9^{\circ} >\phi > 13^{\circ}$) and second sidelobes ($22^{\circ} >\phi > 25^{\circ}$) indicate the region we average over to provide the standard deviations for the noise-to-signal values given in the text. Vertical lines are drawn at $12^{\circ}$ and $25^{\circ}$, corresponding to the radial lines in Figure~\ref{fig:theta_phi_beam}.  The same features present in Figure~\ref{fig:theta_phi_beam} are also clear here: we have a high-significance measurement of the main beam and first sidelobe region, a detection of power in the second sidelobe region, and are dominated by noise where the signal level is small. The data has not been corrected for the beam of the transmitting antenna, and thus the first sidelobe is -0.7\,dB suppressed and the second sidelobe -2.7\,dB suppressed here, as discussed in the text. 
    }
    \label{fig:variance}
\end{figure*}

Figure~\ref{fig:theta_phi_beam} shows a 2D beam pattern in polar coordinates for all four dishes, at frequency 448\,MHz, in a log scale to better depict the low-power sidelobe regions. The beam shapes across all inputs are similar: the maps show a main beam region in the center and two clear sidelobe regions around $12^{\circ}$ and $25^{\circ}$. Most inputs also have a slight excess at the edge of the map at $0^{\circ}$, $90^{\circ}$, $180^{\circ}$, and $270^{\circ}$, with a corresponding excess in the second sidelobe as well. Some features are more apparent in the cartesian gridding (see Figure~\ref{fig:xy_beam} in the Appendix, Section~\ref{sec:appendix}), in particular an asymmetric dip feature between the main beam and first sidelobe present in all inputs. 

As noted in Section~\ref{sec:source}, 1D cut measurements of the bicolog transmitter beam were performed at the NCSU anechoic chamber, but we do not apply corrections from these measurements to the beam results shown here because full 2D data is not available. The measured FWHM of the bicolog transmitter beam from 1D slices at 450\,MHz is 50$^{\circ}$. Because we do not apply a correction for this transmission beam, we artificially suppress the TONE beam maps by 14\% (-0.7\,dB lower) in the first sidelobe at 12$^{\circ}$ and 47\% (-2.8\,dB) at the second sidelobe around 24$^{\circ}$. At most other frequencies in this range, the bicolog transmitter FWHM is $\sim 60^{\circ}$, resulting in an expected 10\% (-0.5\,dB) suppression of the TONE maps at the first sidelobe and 36\% (-2\,dB) at the second sidelobe. This correction is larger than (or at best, the same as) our statistical errors, and should be accounted for in future work when 2D beams will be available. 

Figure~\ref{fig:errors} shows the 1$\sigma$ standard deviation per pixel divided by the pixel value. The average ratio within the FWHM of the main beam at 448\,MHz is 7\% and 9\% for North-polarized and South-polarized inputs respectively. The East-polarized flights have fewer counts, as noted in Section~\ref{sec:gridding}, which drives this difference between the errors. The noise-to-signal is 9\% between $9-13^{\circ}$ (the first sidelobe region), and $\sim70\%$ between $22-26^{\circ}$ (the second sidelobe region). Equivalently, the signal-to-noise ratio is $10-14$ in the main beam and first sidelobe regions, and $\sim$1.4 in the second sidelobe region. Figure~\ref{fig:errors} also shows that regions where the response to the drone signal is very small, namely the sidelobe nulls and locations far from the main beam, are clearly noise-dominated, as expected. 

A slice through the beam for each input and the 1$\sigma$ standard deviation of the same slice is shown in Figure~\ref{fig:variance}. The amplitude of the first sidelobe at $\phi \simeq 12^{\circ}$ is typically about $-15$\,dB and the second sidelobe at $\phi \simeq 25^{\circ}$ is typically around $-25$\,dB. The 1$\sigma$ value is an order of magnitude lower than the signal in the main beam at $\phi \lesssim \pm 4^{\circ}$ and the first sidelobe at $\phi \sim \pm 12^{\circ}$, confirming $\sim$10\% noise-to-signal described from Figure~\ref{fig:errors}. Similarly, the 1$\sigma$ value is also lower than the second sidelobe around $\phi \sim \pm 25^{\circ}$, and roughly meets the signal level in a sidelobe null ($\phi \sim \pm 18^{\circ}$) and far from the main beam beyond $\phi > 28^{\circ}$. We show the evolution of beam properties with frequency for three different slices, in the Appendix in Figure~\ref{fig:freq_phi_beam}. 

\begin{figure*}[t]
    \centering
    \includegraphics[width=1.0\textwidth]{ 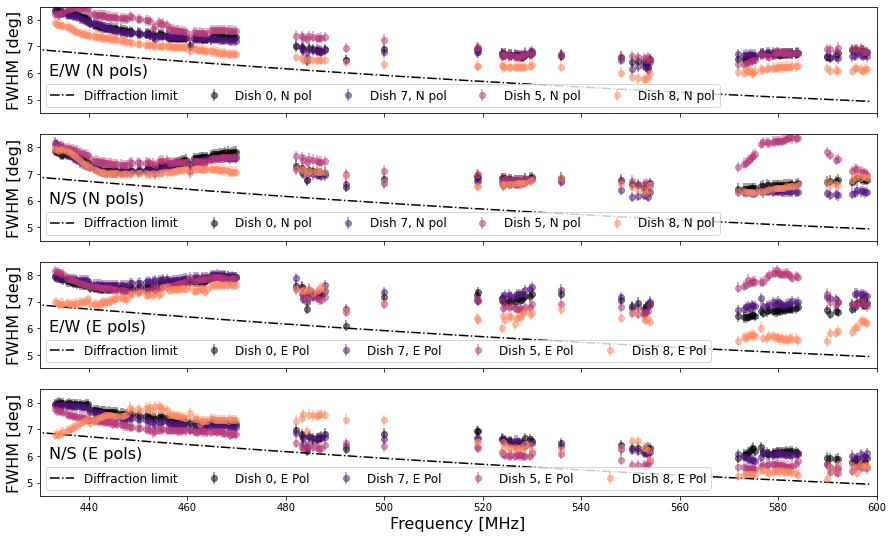}
    \caption[]{FWHM values from coadded maps for all inputs as a function of frequency. (Top) shows the width along the East/West direction for the North-polarized inputs, (Second from Top) shows the width along the North/South direction for the North-polarized inputs, (Third from Top) shows the width in the East/West direction for the East-polarized inputs, and (Bottom) shows the width in the North/South direction for the East-polarized inputs. Also indicated is the diffraction limit for a perfectly illuminated dish, given as 1.03$\frac{c}{\nu D}$ for a $D=6$\,m diameter dish at frequency $\nu$ and speed of light $c$. }
    \label{fig:FWHMs}
\end{figure*}

Finally, (using the process described in Section~\ref{sec:gaussfit}) we fit the FWHM of the main beam using the co-added maps (see Figure~\ref{fig:FWHMs}) obtaining errors of $\pm0.2^{\circ}$. The main beams are mildly elliptical (the E-W and N-S widths are not identical). The widths aligned with the polarization axis are similar between the two polarizations (ie. The N-S width for the North-polarized inputs are similar to the E-W widths for the East-polarized inputs). Beam widths for Dish 5 are clearly higher than for the other inputs at high frequencies; although this was present in the raw data as well, the cause is unknown. Dish 8 (East-Polarized) generally has a lower FWHM, and a dis-similar frequency dependence from the other inputs. This was also seen in the raw data, however we also note that the signal is generally lower for this input (see Figure~\ref{fig:theta_phi_beam}), and so it is possible the gain settings were anomalous. The FWHM values we measure (6$^{\circ}$-8.5$^{\circ}$ across a 450-600\,MHz range) are significantly lower than the values presented in~\citep{TONE_instrument} (7$^{\circ}$-9$^{\circ}$), and we find variation between the two polarizations that does not appear in their results. There are significant differences between the two measurements which may contribute: Sanghavi et al. measured the average FWHM from pairs of two-element cross-product visibilities, while pointing at an off-zenith source (Taurus A), and it was not clear if the dishes had been rotated between the Tau A measurements and the drone measurements we performed. An interesting future measurement would be to tilt the dish and re-measure the beam pattern with the drone to better compare to this celestial source calibration, however this was not attempted during this flight campaign.  


\subsection{Comparison to Simulations}
\label{sec:sims}

The TONE dishes and HIRAX feeds were simulated in CST\footnote{\href{https://www.3ds.com/products/simulia/cst-studio-suite}{Simulia EM simulation suite}}, following the methods described in \citep{saliwanchik21}. The standard HIRAX feed model from \citep{saliwanchik21} was used, which has been validated in antenna range measurements by \citep{Kuhn_HIRAXantenna}. The dish model was modified to match the dimensions of the TONE dishes, which are 6\,m in diameter, with focal ratio of f/D = 0.38. Additionally, four perfect electrical conductor (PEC) cylinders were included to model the ``feed leg’’ support structure between dish and feed. The cylinders were 25\,mm in diameter, forming the edges of a square right pyramid, with the legs meeting at an apex behind the feed, and an apex angle of approximately $45^\circ$. In the simulations performed in \citep{saliwanchik21},  extensive verification was performed showing that the exact value of the conductivity of the metal elements of the model produced no significant difference in the resulting instrument beams at HIRAX operational frequencies. Subsequently, all metal elements have been modeled as PEC for simplicity, and to improve the resource requirements and computational time of simulations. Simulations were performed across the full HIRAX band, from 400-800MHz, in 50MHz increments. Simulations were performed with the CST frequency domain solver, which is faster for wideband simulations and large simulated volumes. Instrument farfield beams were produced with $1^{\circ}$ resolution in $\theta$ and $\phi$.

We compare the simulations at four frequencies to the data at the nearest available frequencies. In Figure~\ref{fig:simsmaps} in the Appendix, we show the simulations on the left, with the data maps on the right, interpolated to the same resolution in polar coordinates. Here we have averaged together all North-polarized inputs and all East-polarized inputs separately, and also averaged adjacent frequency bins to increase the pixel coverage for better comparison. A few things are noteworthy from the maps themselves:
\begin{itemize}

\item There are qualitatively common features at all frequencies between the data and simulations, including the location of the first sidelobe, and pattern of off-axis beam structures. 

\item At 450\,MHz, the simulations predict a more elliptical main beam than we measure. The measured sidelobes are also less well-defined than those found in the simulations, which might suggest that the dish is under-illuminated (this is further supported by full-width-half-max values that are broader than expected). A slight 6-sided structure to the sidelobes is evident in the second sidelobe region, and the quadrupole structure at large angles is also qualitatively consistent with the data.

\item At 500\,MHz, the large angle structure becomes an octopole pattern, this is also qualitatively consistent with the data. 

\item At 600 \,MHz, the East-polarized data seems to show excess power near the main beam on one side. This feature traces directly from an asymmetry unintentionally built into the feed design.  

\end{itemize}

\begin{figure*}[th!]
    \centering
    \includegraphics[width=1.0\textwidth]{ 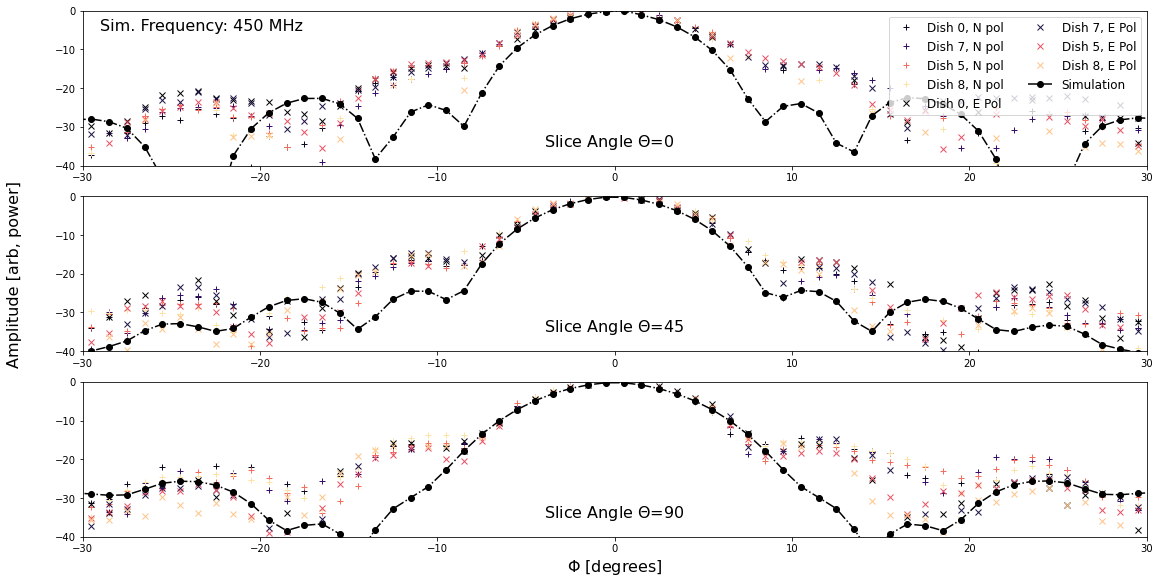}
    \caption[]{One dimensional slice of simulated and measured TONE beams at 450~\,MHz, (Top) for $\theta=0$, (Middle) for $\theta=45$ and (Bottom) for $\theta=90$. $\theta = 0$ corresponds to North. The data from all dishes are shown in color, and the simulation in black. As discussed in Section~\ref{sec:systematics}, we have not included the effects of the transmmitting antenna beam (which would decrease our measurement of the first sidelobes by 0.7dB) nor a geometrical projection effect (which would increase the first sidelobes by 0.2\,dB). }
    \label{fig:simsslices}
\end{figure*}


The asymmetry noted above at 600\,MHz was also measured in in anechoic chamber measurements of the HIRAX antenna alone in \citep{Kuhn_HIRAXantenna}. This was traced back to the feed design, and also generated non-orthogonal linear polarizations at higher frequencies, as will be discussed in Section~\ref{sec:rotate360}. This is a highly significant result, and motivated a redesign of HIRAX feed. The HIRAX collaboration found sub-wavelength physical asymmetries in the HIRAX feed which led to wavelength scale asymmetries in currents propagating on the feed. The redesigned feed has removed the physical asymmetries, and simulations indicate that this redesigned feed should enable HIRAX to significantly reduce asymmetric beam features. This redesign and testing is actively underway and nearing completion.

Figure~\ref{fig:simsslices} compares the beam amplitude of the TONE dishes to simulations in three slices. Shown are 1D cuts through the simulations and the data for all inputs, at 450\,MHz, for $\theta = 0^{\circ}$, $45^{\circ}$, $90^{\circ}$. We find that the sidelobe power is generically higher in the data than the simulations. In addition, the measured beam shapes are quite similar across the three slices, such that a null expected from simulations at $\theta=0, \phi=\pm25$ is not present in the data. We also find the beam patterns are very consistent across dishes and polarizations. 

Discrepancies between simulations and measurements of the feed in isolation were not seen from range measurements \citep{Kuhn_HIRAXantenna}, and thus the most likely explanation is that the discrepancies originate in the dish system itself.  For example, they may be the result of deformations in the dish surface, such as those due to gravity, which are not modeled in our current simulations. Additional simulations are currently underway that incorporate surface deformation into existing dish models to investigate the effect these may have on instrument beams: \citep{gerodias23} models surface deformations in the HIRAX dish and demonstrates that they can lead to increased sidelobe amplitude and shifts in sidelobe angular positions.

Finally, the FWHM from the data is clearly wider than the simulations predict. We find, for example, at 450\,MHz the simulation predicts an elliptical main beam with X-width of $6.7^{\circ}$, while we find values of $7.1-8^{\circ}$ and similarly for Y-widths the simulations predict $7.7^{\circ}$ while we find values of $7-7.4^{\circ}$. The diffraction limit would predict $6.5^{\circ}$. The observed discrepancies in beam-width may be caused by under-illumination of the dish, possibly due to variations in feed location which shift the feed away from the phase center of the feed-dish system. Updated versions of the HIRAX dishes include feed mounting mechanisms which more reliably fix the feed location to the phase center. 


Broadly, disagreements between simulations and measurements underscore the importance of performing detailed calibration measurements for observatories, and that relying solely on beam simulations during analysis may be insufficient.



\subsection{360 Degree Rotations}
\label{sec:rotate360}

\begin{figure*}[t]
    \centering
    \includegraphics[width=1.0\textwidth]{ 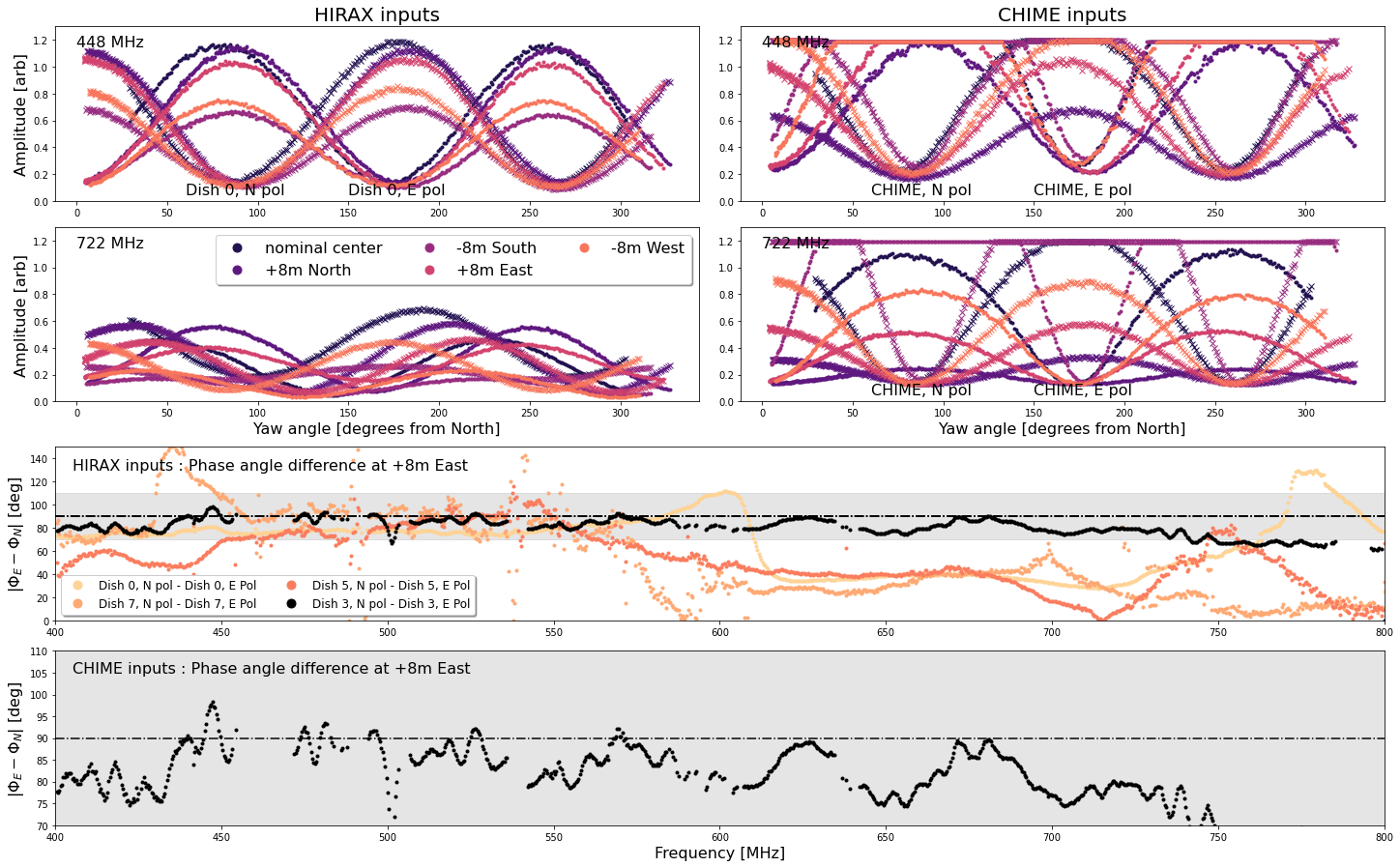}
    \caption[]{
    (Upper left) Signal amplitude vs yaw angle at 448\,MHz for Dish 0, populated with a HIRAX-style feed. The colors indicate location of the 360$^{\circ}$ rotation (nominal center, and four locations corresponding to 8\,m distances in the four cardinal directions, as described in the text). The markers indicate E-polarized and N-polarized inputs. In this panel, the two polarizations are clearly separated by $\sim90^{\circ}$ for all locations in the beam. (Upper right) The same frequency, markers, and colors as the left panel, but for Dish 6, instrumented with a CHIME-style feed. The CHIME data is clearly saturated (a flat line) for most locations, but the $\sim90^{\circ}$ difference betwee polarizations is still apparent for all locations. (Middle left and Middle right) The same as above, but for frequency 772\,MHz. The CHIME feed continues to show an orthogonal pattern between the two polarizations, however the HIRAX input shows that the two polarizations are no longer orthogonal, which is true for all locations. (Center) Phase difference between the E-polarization and N-polarization inputs for three dishes populated with HIRAX feeds, for the location +8m East. At low frequencies below 600\,MHz, the phase difference is $90^{\circ}$, as expected, but diverges at high frequencies. The grey band indicates $\pm20^{\circ}$. (Lower) The phase difference for Dish 6, instrumented with the CHIME feed. Here, the difference is consistently near $90^{\circ}$ for all frequencies, as expected. The range in this plot is also shown as the grey band in the HIRAX-style feeds in the plot above. The cause of the high-frequency (10\,MHz) ripple is unknown.    
    }
    \label{fig:rotations}
\end{figure*}

To measure the polarization angles of the dishes, we performed a special rotation flight in October when the RTK unit was not functional. The drone flew to the nominal center of the Dish 0 and rotated in yaw by 360$^{\circ}$ over 19 seconds. We then re-positioned the drone 8\,m North and repeated the rotation, and repeated the rotations 8\,m in the South, West, and East directions, for a total of five positions. The power measured during the rotations should peak when the drone transmitting antenna polarization and telescope polarization are aligned, and be minimal when the two polarizations are anti-aligned. 


As the drone rotates, the polarization angle of the broadcast calibration signal also rotates. Consequently, the measured amplitude of the calibration signal should trace a sinusoid. As was true in  Section~\ref{sec:processing}, the drone and telescope timestamps are not initially synchronized. Because this was not a gridded flight, we could not use the Gaussian fitting outlined in Section~\ref{sec:processing} to determine the correct time offset. Instead, we found that when the drone rotated more than $360^{\circ}$, an offset in timestamp would generate two different amplitudes at the same reported yaw angle. This is because when the drone reports $0^{\circ}$ and then (some time later) 360$^{\circ}$, the correlator time stamp is reporting a time which corresponds to two different angles. We correct the time stamps by shifting the drone timestamp until the telescope data matches at all angles. We found an offset of 3.7\,s resulted in good alignment, which is large but within the range of gridded flight time offsets discussed in Section~\ref{sec:processing}.

The signal amplitude as a function of yaw angle after this correction is shown in the upper four panels of Fig~\ref{fig:rotations}, where the different colors indicate the different locations where the drone performed a 360$^{\circ}$ rotation. The panels are divided by dish (Dish 0, populated with a HIRAX feed, and Dish 6, populated with a CHIME feed) and frequency (448\,MHz and 722\,MHz), and both linear polarizations are shown in a given panel. Note, in this section, we are using the full 400-800\,MHz band to compare polarization angles across the entire TONE frequency range, and we did not use data from Dish 6 in the preceding sections. We find that the linear polarizations are essentially orthogonal for dishes with either HIRAX or CHIME feeds at low frequencies. However at higher frequencies, the two inputs of the the HIRAX feed are no longer orthogonal even though the two CHIME inputs remain orthogonal. To quantify this behavior, we fit this data with the function 
$$
S = A~\mathrm{cos}^{2}[P (\mathrm{Yaw} - \delta)] + C
$$
where $S$ is the measured signal, $A$ is the amplitude, $P$ is the period, $\delta$ is the phase, and $C$ is an amplitude offset. This is the expected functional form for the intensity of polarized light passing through a polarized filter. A well-behaved polarization response should have a period ($P$) of 180$^{\circ}$ and a phase ($\delta$) of either 0$^{\circ}$ (for North-polarized inputs) or 90$^{\circ}$ (for East-polarized inputs) if the telescope and drone polarizations are perfectly aligned to GPS north, and thus a phase difference of $90^{\circ}$ between the two orthogonal polarizations.  

We chose the +8m East location for the fitting results presented in the remaining panels of Figure~\ref{fig:rotations}. Although the largest signal comes from the `nominal center' location where it is nearest to on-axis and thus has the most well-behaved polarization response, this location had poor fits in the CHIME data due to saturation. We chose the +8m East because it had good signal to noise for all inputs considered and the phase and period for the CHIME inputs were well fit. As a result, we expect this location to be at least a few degrees away from the center of the beam for both Dish 0 and Dish 6, and thus some additional cross-polarized signal will contribute signal. Despite this, we note the overall behavior is not dependent on location: the signals are nearly orthogonal at low frequencies for all locations in the beam, and clearly non-orthogonal for the high frequencies for the HIRAX input.  At low frequencies where all inputs are well behaved, the phase for the North-polarized feeds was typically $-3 - 10^{\circ}$ and for the East-polarized feeds was typically $88-90^{\circ}$, indicating that the measured polarization axes are roughly aligned to the cardinal directions, but are not perfectly orthogonal. Because the phases are dish-dependent, likely some of the rotation is due to dish and feed polarization orientation, as well as our choice to use a location at least few degrees off-axis. 

We show the difference in phase angle between the two polarized inputs for the pairs on the HIRAX dishes (lower) and CHIME (lowest) panels of Figure~\ref{fig:rotations}. At low frequencies, all feeds show a phase difference near $90^{\circ}$, as expected. The CHIME feed shows a difference of $\sim 90^{\circ}$ across all frequencies, with a slow decline to higher frequencies. This is likely due to our choice to use a position a few degrees from boresight. A small frequency-dependent ripple is present in this difference, the cause is unknown. At high frequencies above 600\,MHz the HIRAX feeds become less orthogonal, with a phase difference of $\sim$50 degrees for all three dishes considered. Eventually, the cause of this polarization offset was traced to an asymmetry in the design of the HIRAX antenna as described in Section~\ref{sec:sims}.  Ultimately this drove our decision to consider only frequencies below 600\,MHz in this paper. 


\subsection{Cross-polarized beam properties}
\label{sec:polarized}

\begin{figure*}[t]
    \centering
    \includegraphics[width=0.9\textwidth]{ 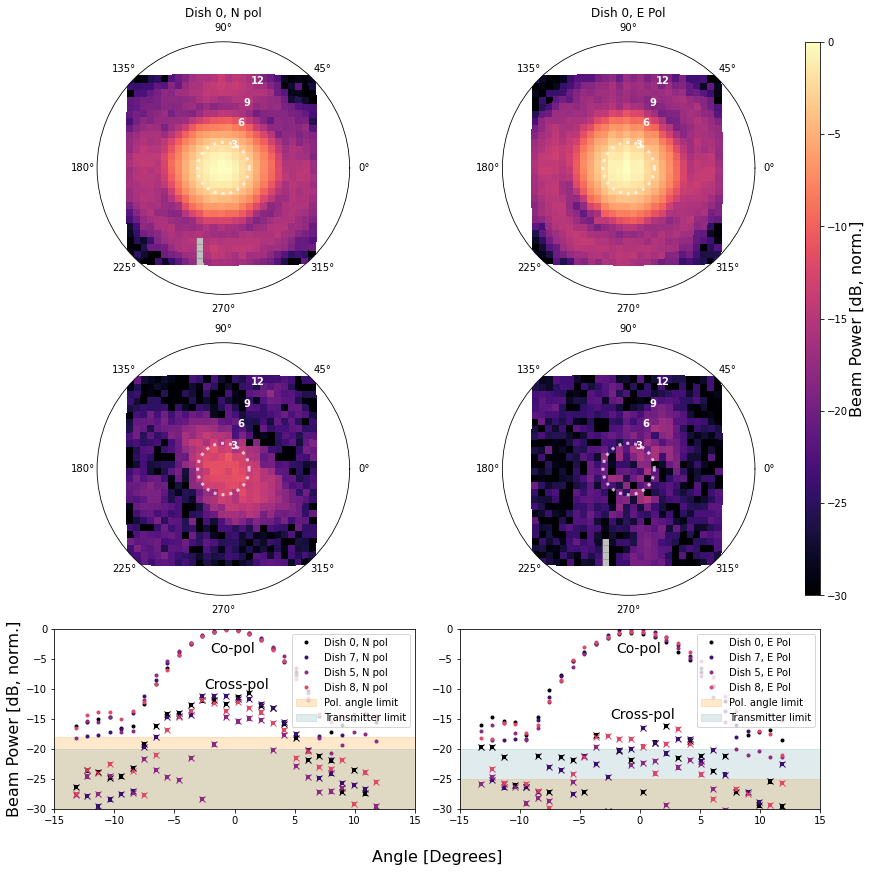}
    \caption[]{(Upper left) Map of Dish 0, North-polarized input in co-polarization, for flight 618 at frequency 448\,MHz. The white circle denotes the approximate FWHM of 7.5$^{\circ}$. (Upper right) Co-polarized map of Dish 0, E-polarized input for flight 620. (Center left) Cross-polarized map, with normalization discussed in the text, for Dish 0, North-polarized input. (Center right) Cross-polarized map for Dish 0 East-polarized input. (Lower left) A slice through the center of the beam, now showing North-polarized inputs from all dishes. The co-polarized map is shown as circular markers, and the cross-polarized are crosses. The level of cross-polarized is roughly similar across inputs, and the $\pm8$\% errorbar (as described in the text) is present for the cross-polarized values. Shaded regions show limits from systematics as discussed in the text, for contributions from mis-aligned polarization angle and the transmitter cross-pol.  (Lower right) Same as lower left but for for East-polarized inputs. The North-polarized inputs are clearly not systematics limited, however the East-polarized cross-polarized is low enough that the -20\,dB cross-polarized amplitude from the transmitting antenna may be limiting our measurement. 
    }
    \label{fig:N_crosspol}
\end{figure*}


During a flight, we measure the signal from both linear polarizations on a dish: the polarization nominally aligned with the drone's orientation (co-pol) and the polarization orthogonal to the drone's orientation (cross-pol). We used our two reference flights (the North-polarized flight 618 and the East-polarized flight 620) to investigate the cross-polarized pattern of the dishes. Both reference flights are main beam flights with the same transmit power level and form a dense grid in the main beam and first sidelobe region.

Flights 618 and 620 comprise a set of co- and cross-polarized measurements for all feeds. The normalization from the co-polarized flight for a given feed is applied to both the co- and cross-polarized data for that feed. The two flights were performed at slightly different heights (174\,m for flight 618, 178\,m for flight 620), which artificially suppressed the cross-polarized amplitude for the North-polarized inputs and artificially increased the cross-polarized amplitude of the East-polarized inputs by 4\%. We correct for this shift, which changes the cross-polarized values by $\sim$0.7\,dB. The cross-polarized map was then formed using the gridding procedure described in Section~\ref{sec:gridding} onto a cartesian grid and finally projected into polar coordinates. The cross-polar maps and co-polar maps for both polarizations of Dish 0 are shown in Figure~\ref{fig:N_crosspol}, with slices shown in the panels underneath. We find cross-polarized values in the main beam range between $-23.5$\,dB to $-12.5$\,dB relative to co-polarized peak amplitude, and find no clear `dip' in the center as would be expected from a cross-polarized data set, which we attribute to imperfect angle alignment during the flights.

The cross-polarized beam structure appears elongated in one direction (roughly $-45^{\circ}$ for the North-polarized input and $+45^{\circ}$ for the East-polarized input). This direction was common to all dishes and polarizations and so it is possibly associated with dish structure (eg. feed leg locations). The cross-polarized amplitude within the main beam was consistently higher for North-polarized inputs than for East-polarized inputs. Dish 0, East-polarized input had a cross-polarized amplitude of $-21.5$\,dB ($\sim$1\%), with other inputs ranging between $-18$ to $-23$\,dB. In comparison, the Dish 0 North-polarized input has a cross-polarized peak of $-12.5$\,dB ($\sim$5\%), with other inputs ranging from $-12$ to $-16$\,dB. The simulations assumed both polarizations were identical, and so would not have recovered polarization-specific cross-polarized patterns. Below, we detail some factors which may contribute to the discrepancy between North- and East-polarized cross-polarized amplitudes. 

We investigated the effect of gain variation on this measurement, as follows. In the normalization scheme described above, gains from the telescope, changes in the source brightness, and changes in noise could have contributed to errors in our estimates of the cross-polarized value. To investigate gain changes, we examined the best-fit Gaussian amplitude for the Dish 0, North-polarized input for flight 618 and 623 (flight 623 was the next available flight after 620). We found the Gaussian amplitude increased by 4\% at frequency 448\,MHz. To be slightly more conservative, we took $\pm$8\%, twice the gain increase, corresponding to the largest variation we found in spectrum analyzer measurements (as discussed in Section~\ref{sec:instrument}) and similar to the average statistical error in the main beam (as discussed in Section~\ref{sec:mainbeam}). This corresponds to an error bar of 0.7\,dB, which is not enough to explain the difference in cross-polarized amplitudes between the polarizations, but which we include as an error bar on the cross-polarized values in Figure~\ref{fig:N_crosspol}.

In addition, the drone's polarization axis is not perfectly aligned to the feed polarization angle, as discussed in Section~\ref{sec:rotate360}. We did not necessarily measure the polarization angle at the beam center, but for Dish 0 and frequency 448\,MHz, the polarization angles with good fits and the highest signal were $1-3^{\circ}$ for the East-polarized feed (depending on location) and $7^{\circ}$ rotated for the North-polarized feed. This produces a floor of around $-18$\,dB\footnote{cross-polar leakage in dB is $20 \mathrm{log}_{10}(\mathrm{sin}(\theta))$} for the East-polarized inputs and $-25$\,dB for the North-polarized inputs from a contribution of co-polar power leaking into the cross-polarized purely from the rotation, also indicated in Figure~\ref{fig:N_crosspol}. We should emphasize this is likely to be a conservative estimate since we would expect the angles to have better alignment than 1-7$^{\circ}$ if these rotation locations came from the center of the beam. The measured cross-polarized signal in both cases is larger than this floor, indicating the results from the main beam are robust to location and the difference between the North-polarized and East-polarized cross-polarized signals are not significantly dependent on dish, thus we conclude this asymmetry is a real effect of the telescope system. 


Finally, the drone transmitting antenna has its own cross-polarization pattern, which could effect our measurement of the cross-polar pattern of the TONE dishes. Ideally this could be be removed from the TONE cross-polar beam pattern, however we have measured only 1D cuts through the transmitting antenna beam and do not have 2D information (see discussion in Section~\ref{sec:source}). However, we can still use the 1D cuts to assess whether it is likely that the bicolog cross-polar pattern is a significant contributing factor to the TONE measurement. In the 1D cut measurements, the cross-polarized peak amplitude in both 1D cut directions were \aprx$-20$\,dB below the co-polar peak amplitude. This is smaller than the level of cross-polarized in the North-polarized drone measurements of TONE, thus the excess is unlikely to be caused by the cross-polarized pattern of the transmitting antenna. However, -20\,dB is roughly the same level as the East-polarized cross-polarized peak, thus could be limiting our measurement of the TONE cross-pol. 

\subsection{Summary of Errors and Systematics}
\label{sec:systematics}

In this section, we summarize the statistical and systematic errors outlined in this paper, which appear in Table~\ref{tab:errors}. First, as described in Section~\ref{sec:mainbeam}, our statistical errors within the FWHM of the beam are 7-9\% (depending on polarization), 9\% at the first sidelobe, and 70\% at the second sidelobe. Because our measurement is consistent with the Ludwig-I polarization convention, we overpredict the sidelobe levels by 4\% (0.2\,dB) at the first sidelobe and 20\% (1\,dB) at the second sidelobe (relative to Ludwig-III). This is less than our statistical errors, but using a flight pattern and payload that would be consistent with the Ludwig-III convention would be required for a 1\% beam measurement. Because we do not have 2D beam information (see Section~\ref{sec:source}) from the drone transmitting antenna we are unable to compensate for its transmission properties and as a result we underestimate the sidelobe power by $\sim$ 14\% (0.7\,dB) in the first sidelobe and 47\% (2.8\,dB) in the second sidelobe, an effect which can be larger than our statistical errors (discussed in Section~\ref{sec:mainbeam}). Finally, as discussed in Section ~\ref{sec:polarized}, the -20\,dB cross-polarized peak amplitude from the transmitting antenna is subdominant to the cross-polarized peak amplitude we measure in the North-polarized inputs, but at a similar level to the cross-polarized peak amplitude we measure in the East-polarized inputs. 

\begin{table}[ht]
\centering
\begin{tabular}{|l|l|l|l|}
\hline
Source & Main Beam & 1st Sidelobe  & 2nd Sidelobe  \\
& & (12$^{\circ}$) & (24$^{\circ}$) \\ \hline
Statistical & 7-9\% & 9\% & 70\%  \\
Trans. beam  & - & 14\% & 47\% \\
Trans. beam (pol) & -20\,dB & -20\,dB & -20\,dB  \\
Ludwig-I  & - & 4\% & 20\% \\ \hline
\end{tabular}\\
\caption{\label{tab:errors} Summary of systematic and statistical errors, more detail is provided in the text. Worst-cases have been represented here to be conservative. 
}
\end{table}

\section{RFI measurements}
\label{sec:rfi}

The drone flights at GBO were conducted during maintenance periods when radiofrequency interference (RFI) was unrestricted. However, future dishes which may benefit from drone calibration are likely to have more stringent RFI limits. CHIME, a transit radio interferometer at the Dominion Radio Astrophysical Observatory, operates in a radio quiet zone. Therefore, flights must be coordinated to occur during maintenance periods to prevent contaminating the data acquired by other experiments. HIRAX will be constructed at the SKA SA (Square Kilometer Array South Africa) site, approximately 2.3\,km from the nearest SKA dish, thus any RFI generated by the drone below 6\,GHz is regulated. To assess our broadband RFI levels and compare them to existing regulations, we performed a series of drone RFI measurements at Green Bank Observatory. We categorize our RFI sources as either intentional or unintentional emitters. intentional emitters include known sources of RFI (our own radio source and communications channels) while unintentional emitters are sources of RFI that would not appear in the drone specification sheet (e.g. from the motor rotors, switching power supplies, etc).

The RFI limits for the SKA site are divided into two groups: those for MeerKAT and those for SKAMID. RFI emission limits are defined in units of EIRP (equivalent isotropic radiated power at the transmitter) and the limits are calculated using the methodology described in SKA document SSA-0008A-038 \textit{Applying Telescope Protection Levels}. A path loss factor is added based on calculations described in SKA document SSA-0008J-108 \textit{HIRAX RF Attenuation Report}. In EIRP units, the SKA limits range from $-90$ - $-70$\,dBm across the band (0-6\,GHz) while the MeerKAT limits range from $-60 - 0$\,dBm for the same frequency range. In all cases, the measured RFI emission from the drone (and additional digital instrumentation required for drone flights) exceeds the SKA limits. We therefore plot only the MeerKAT limits in the following sections. 

 

\subsection{Unintentional Radiators}

\begin{figure}
    \centering
    \includegraphics[width=1.0\textwidth]{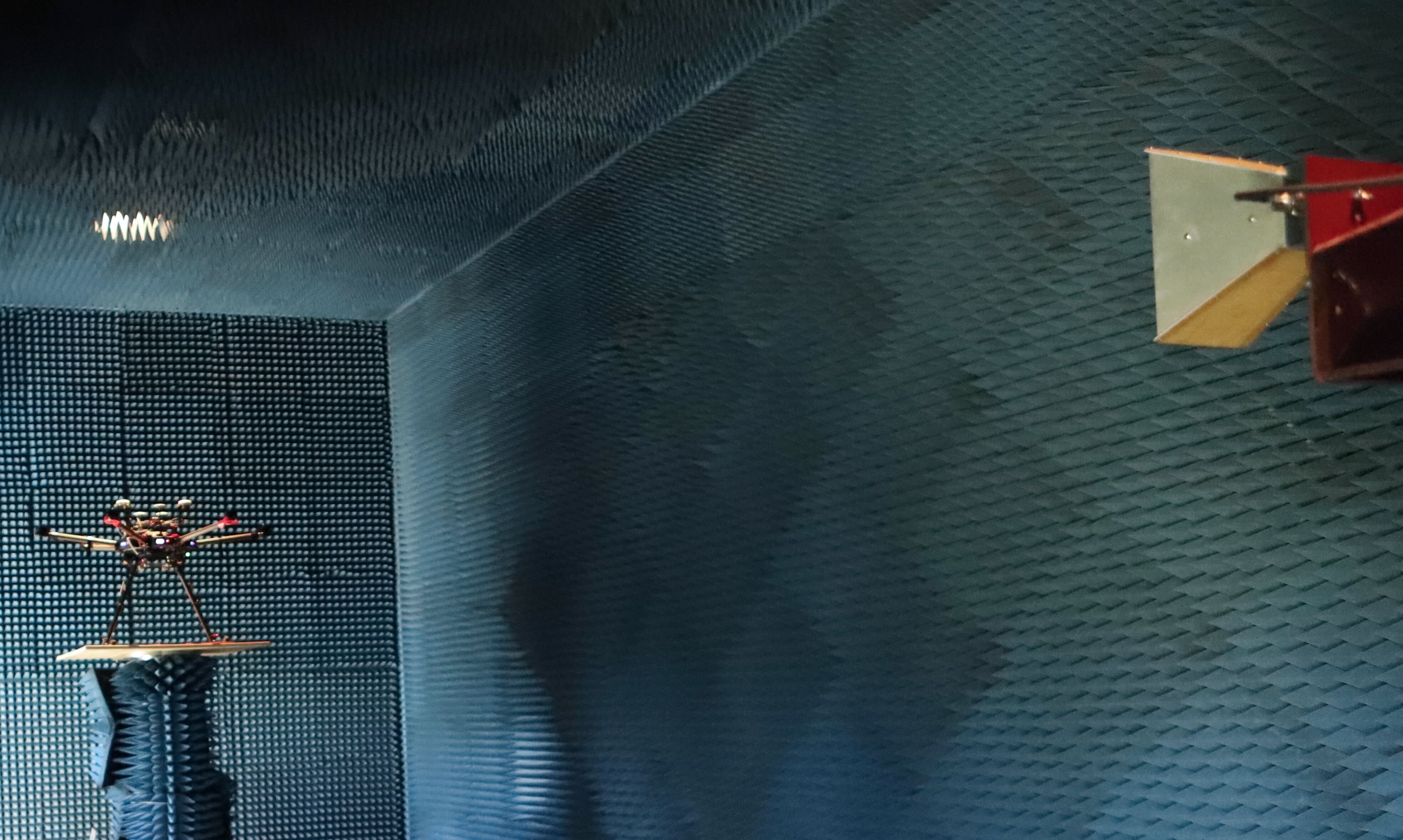}
    \caption{Experimental measurement setup inside GBO anechoic chamber. In foreground is a standard gain horn, in background is the drone and payload on a custom non-reflective wooden mount.}
    \label{fig:rfi_setup}
\end{figure}

The drone contains high-speed digital communications, DC-DC converters, and motors, all of which are expected to produce RFI. To measure this RFI, we used the anechoic chamber at GBO (see Figure \ref{fig:rfi_setup}), which is typically used for certifying devices and enclosures to be placed near the Green Bank Telescope.


Green Bank's anechoic chamber measures 15 feet by 15 feet by 37 feet, and is lined with AEP-6 and AEP-8 absorber foam. It is equipped with a variety of antennas and receivers ranging from 100MHz up to 115GHz \footnote{\href{https://www.gb.nrao.edu/electronics/tool_index.shtml}{GBO measurement procedure}}. Working with GBO staff, we followed the standard measurement procedure to calibrate the instrumentation and then measure the emission from the drone. The drone was placed on its landing legs and measurements were taken at each of 4 azimuth angles. Then, the drone was rotated 90 degrees and placed into a custom RF-transparent mount so that the bottom of the drone pointed towards the measurement antenna, equivalent to what would be transmitted to the telescope being measured if the drone were flying overhead. Finally, the drone and mount were rotated through 12 azimuth angles, simulating different degrees of tilt of the drone over a telescope. We did not measure each piece of digital equipment separately, so here we report the aggregate emission from all drone-related digital equipment.

\begin{figure}
    \centering
    \includegraphics[width=1.0\textwidth]{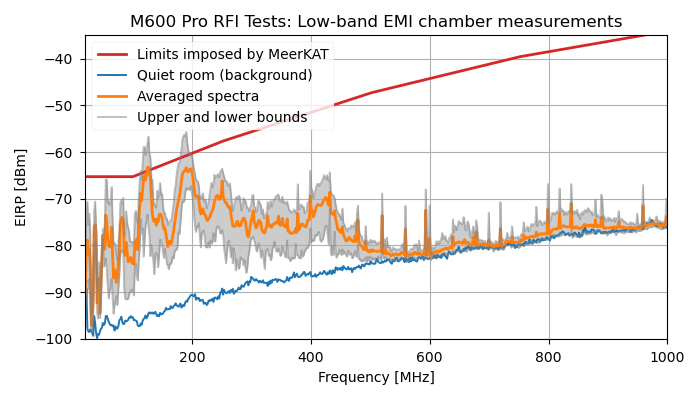}
    \caption{Unintended drone RFI emission as measured in the anechoic chamber at GBO. Total unintended emission meets required limits from MeerKAT (shown in red) except in two regions, where the upper bounds exceeded the requirements.}
    \label{fig:rfi_chamber}
\end{figure}

The measured RFI spectra are shown in Figure \ref{fig:rfi_chamber}, where the lines plotted show the maximum, minimum, and median values across all angles. There is a clear excess of RFI above the background level at frequencies below 600\,MHz. The range of values shown for different angles is larger than the measurement errors, indicating that emissions from the drone may vary by up to 10dB depending on the angle. On average, the RFI from unintentional emitters were within the limits set by MeerKAT, except for at approximately 170MHz, where they exceed the limit by $<$2\,dB. In the worst-case orientation, the unintentional radiators exceeded the limits imposed by MeerKAT by approximately 5\,dB in two regions. This low-frequency excess is also seen in measurements acquired with TONE while the drone was flying over the telescope, Figure~\ref{fig:rfi_on_off}.

In addition, we tested the drone's brushless motors to confirm that they were not a source of low frequency RFI.  It was not possible to fly the drone within the anechoic chamber, so a laptop running DJI Assistant 2 was used to spin up the motors without the rotors attached. We found that the measurements with the spinning motors and laptop were indistinguishable from the measurements with the laptop and stationary motors, thus the motors are not contributing a significant amount of RFI above the ambient RFI from the drone and other components.

\subsection{Intentional Radiators}
In addition to the noise source payload, the system requires several other intentional radiators - devices whose purpose is to transmit RF signals to assist in the drone's functioning.
All of our intentional radiators have passed FCC certification and have test documents available.
The tests were performed in the same chamber described above, using a specialty calibrated Electromagnetic Compatibility (EMC) receiver.
The intentional radiators in our drone system are summarized in Table \ref{tab:rfi_table}. 
Included in Table \ref{tab:rfi_table} is the RFI brightness above or below the requirement from MeerKat.
From this table, there are three sources which are significantly brighter than the requirements: the drone controller at 2.45 GHz, the drone telemetry link also at 2.45 GHz, and the RTK air unit at 900 MHz. 

The intentional emitters are by far the strongest source of RFI on the drone, as expected. Using maintenance windows may ultimately be a possibility, but mitigation strategies would be wise to consider. The RFI requirements are far less stringent at higher frequencies, and thus if communications can be moved to higher frequencies it may bring drone emissions into compliance. New RTK units now use 5\,GHz for communications, and although they are significantly brighter than the 900\,MHz transmitters, the RTK ground unit could be shielded from telescopes because it sits on the ground. Newer drones also use 5\,GHz for drone controller and telemetry communications, and may therefore meet MeerKAT requirements. 

\begin{table}[h]
    \centering
    \begin{tabular}{l|l|l|l}
        \textbf{Source} & \textbf{Frequency} & \textbf{Power} & \textbf{dB above}\vspace{-0.5em} \\
         & & \textbf{(EIRP)} & \textbf{MeerKAT limit}\\ \hline \hline
        Drone controller & 2.45 GHz & 5.46 dBm & 21.63 dB \vspace{-0.3em}\\
        (low band) & & & \\\hline
        Drone controller & 5.8 GHz & 0.09 dBm & -4.85 dB \vspace{-0.3em}\\ 
        (high band) & & & \\\hline
        Drone telemetry & 2.45 GHz & 2.25 dBm & 18.41 dB \\ \hline
        RTK air unit & 915 MHz & 9.66 dBm & 45.65 dB \\ \hline
        RTK ground unit & 915 MHz & 9.16 dBm & 45.15 dB \\ \hline
        Noise source & 400 MHz -& $\sim$-65 dBm & $\sim$-20 dB\vspace{-0.3em}\\ 
        (43dB atten.) & 800 MHz & & \\\hline
    \end{tabular}
    \caption{Intentional Radiators on the DJI M600}
    \label{tab:rfi_table}
\end{table}

\subsection{RFI with Telescope Data}

\begin{figure}
     \centering
     \includegraphics[width=1.0\textwidth]{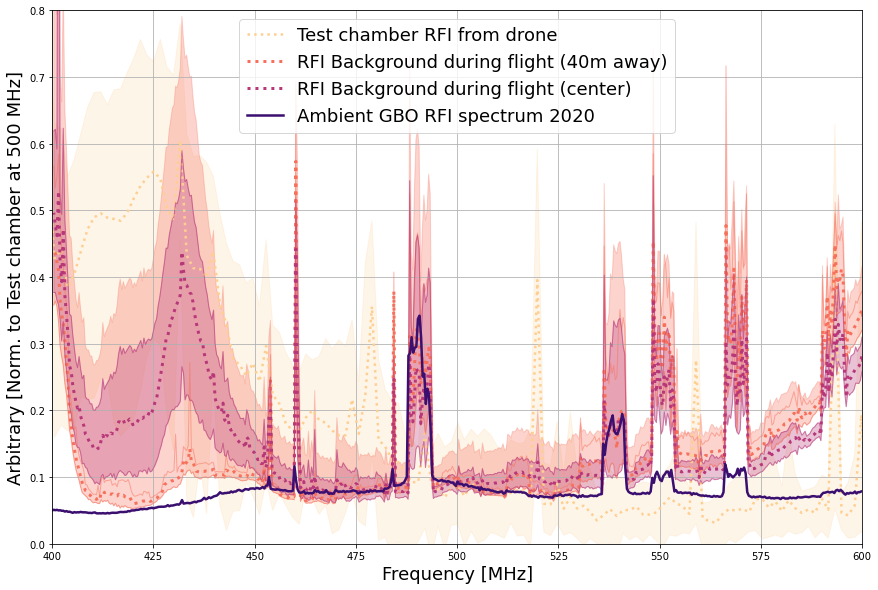}
     \caption{Drone RFI measurements at Green Bank Observatory. Shown in this figure are spectra acquired by the TONE telescope while the drone (with onboard calibration source off) is within 5$^{\circ}$ of zenith pointing per dish (rose line)  and the average maximum and minimum value (rose band). RFI from the drone measured in test chamber with no external RFI is shown in yellow, with the yellow line indicating average over many positions and yellow band shows max and min of those positions. Finally, the spectrum measured through the TONE antenna shows the ambient RFI at GBO (black). The amplitude is arbitrary, and values have been scaled to show them on the same plot. The emission from the drone while the RF source is off can be accounted for by the combination of ambient RFI and known RFI from the chamber measurements.}
     \label{fig:rfi_on_off}
 \end{figure}

As described in Section~\ref{sec:processing}, we separate data points where the drone signal is on from when the drone signal is off, and difference, to form a background-subtracted signal data set. Most of the analysis presented in this paper used the `drone on' data, however the `drone off' data can be used to estimate the RFI power from the drone itself. We selected one flight, flight 625, and chose only the data points where the signal was in the `off' state. We isolated when the drone was directly overhead (essentially over Dish 0), when the drone was 40\,m away, and 80\,m away from that central point. The minimum, maximum, and mean value at each position was calculated per dish, and the average across dishes was computed. Figure~\ref{fig:rfi_on_off} shows these values for two locations: directly above the dish and 56m so the SouthWest (40m to the West, 40m to the South) from dish center. Also shown for comparison are an ambient RFI spectrum measured at GBO during the flight campaign, and a drone RFI spectrum measured in the GBO anechoic chamber. The drone flight and ambient RFI spectra are scaled to match the chamber measurements at 500\,MHz, although we have not accounted for the telescope gain or other factors that could significantly influence the frequency dependence of these spectra. As a result, only broad similarities between the measurements can be discussed. We also limit ourselves to the same 400-600\,MHz range we have considered in this paper, mainly because the shape of the background above 600\,MHz appeared to be dominated by digital gain settings, which we confirmed by comparing to a measurement conducted while the drone was not flying.  We find that all measurements support excess emission from the drone at low frequencies, with the most significant emission of RFI above the background level occurring between 400-450\,MHz. This corresponds to the beginning of the frequencies which passed data selection, so the drone RFI was likely limiting our measurement abilities at the lower end of the band. The two spikes above 450\,MHz and the top-hat shaped excesses around 490\,MHz, 540\,MHz, 550\,MHz, and 570\,MHz are human-generated RFI that are not caused by the drone, a claim supported by their appearance in the ambient GBO RFI spectrum. There is one remaining excess (at 590\,GHz) which appears in the drone data, that is faintly present in the GBO ambient spectrum, which we also attribute to human-generated site-based RFI. 

\section{Summary}
\label{sec:summary}

This paper presents drone-based measurements of four dishes within the TONE array between frequencies 400-800\,MHz. We described the drone, payload instrumentation, flights, and data processing required to make and co-add beam maps. We presented co-added maps in co- and cross-polarization, data from 360$^{\circ}$ rotation flights for polarization alignment, and measurements of the RFI spectrum from the drone in an anechoic chamber. 

Using drone-based beam maps, we found that the co-polarized beam shape was similar across the four TONE dishes, each having a first sidelobe around $12^{\circ}$ and $-20$\,dB down from the peak center, and a second sidelobe around $25^{\circ}$ and $-25$\,dB down from the peak center. The main beam signal was measured with statistical errors of 7-9\%, possibly with a significant contribution from calibration-source variability. The first and second sidelobes were measured with statistical errors of 9\% and 70\%, respectively. The full-width-half-max values (FWHMs) decrease with frequency from 8.5$^{\circ}$-6$^{\circ}$ with significant differences between polarizations and dishes. On-sky measurements from published TONE interferometric data found somewhat larger FWHM values and no clear differences between polarizations and dishes~\citep{TONE_instrument}; the cause of this difference is not currently understood. The beams were qualitatively similar to simulations, however we found generally higher sidelobe levels and discrepancies in location for higher-order sidelobes. These effects are likely due to dish features that are not modeled in our simulations, such as surface roughness and large scale deformations (e.g. due to gravity). \citep{gerodias23} shows that dish surface deformations can lead to higher sidelobe amplitudes and shifts in the locations of sidelobes.

Using data from 360$^{\circ}$ rotation flights to assess polarization alignment, we determined that the polarization axes of the reflectors with both CHIME and HIRAX feeds were orthogonal at low frequencies. For reflectors with HIRAX antennas, at higher frequencies the polarization axes deviated from orthogonality, which spurred further investigation that tied this effect back to the HIRAX feed design~\citep{saliwanchik21}. This motivated presenting only beam maps below 600\,MHz in this paper to avoid frequency-dependent cross-polarization signals. In the lower end of the band where the polarization axes are well defined, we found cross-polarized levels of $-20$\,dB for the East-polarized inputs, roughly the same level as expected from the drone transmitter's own cross-polarization. We found cross-polarized levels of $-15$\,dB for the North-polarized inputs, a 5\,dB difference between the two polarizations. This difference was too large to be explained by either imperfect alignment between the drone and telescope polarization axes or known gain variations, and thus may be indicating a truly higher cross-polarization pickup in one polarization. 

We found that RFI measured during flights was consistent with a combination of RFI emission from the drone (which was independently measured in an anechoic chamber) and human-generated RFI sources around Green Bank Observatory. The drone and its peripherals (e.g. laptop, controller) emit measurable unintended RFI  below 550\,MHz. Although most emissions are still within common RF quiet site requirements, continued efforts to mitigate RFI are warranted, particularly below 430\,MHz.

We also identified limitations in the current data sets and analysis methods that could be improved with new flight trajectories, updates to the payload, and improved analysis techniques. First, we did not correct for the beam pattern of the transmitting antenna used to broadcast our calibration signal from the drone because we only had 1D cuts instead of full 2D maps. From existing 1D measurements, we estimate that this is our largest systematic, causing a bias larger than our statistical errors in the first sidelobe and larger than the precision requirement of 1\% everywhere.  Second, our planar grid measurement with the transmitting antenna fixed to the drone chassis provides a Ludwig-I convention polarization measurement. Making a measurement with the standard Ludwig-III convention will require further updates, in particular a gimbal mount~\citep{10336719}. This is worthwhile to consider, as it was our second largest source of systematic error (but smaller than statistical errors in this paper). We would also benefit from a more flexible polar gridding scheme, which reduced the number of well-measured pixels near the center of the beam. Finally, gain variations of the calibration source appear to be one of the primary drivers of our uncertainty in the final co-added maps; confirming the source's stability on hour time scales is essential. 


These measurements represent many firsts: 

\begin{itemize}
\item The first time a drone was used to compare the beam shapes across multiple reflector dishes. 
\item The first time a cross-polarized 2D beam map has been published. 
\item The first time a switching calibrator source was implemented across a broad band. This switching functionality allowed us to difference backgrounds from the calibrated transmission. This background would have dominated our errors if not removed. 
\item The first investigation of polarization angle alignment using drones, which found design flaws in the HIRAX feed at frequencies above 600\,MHz. 
\end{itemize}

\section{Acknowledgments}

We gratefully acknowledge the assistance from the GBO staff during the measurement campaign, and use of facilities such as the anechoic chamber and antenna range. We particularly thank the TONE array team, in particular Pranav Sanghavi and Kevin Bandura, for help with the array during those measurements. We also thank Danny Jacobs and Cynthia Chang for helpful discussions on analysis and technological approaches. ERK thanks Mike Seiffert for helpful conversations and feedback during the writing process. We depended on Yale Wright Lab machine shop and support staff, as well as the YCRC computing center for simulation support. Work for this paper for BRS, ERK, LBN, and WT was supported by the National Science Foundation, Grant No. 1751763. We also thank many undergraduate students who contributed to the analysis presented here: Ry Walker, Audrey Cesene, Spencer Greenfield, Jordan Davidson, and Eli Bader. ERK was supported by a NASA Space Technology Research Fellowship and an appointment to the NASA Postdoctoral Program at the Jet Propulsion Laboratory/California Institute of Technology, administered by Oak Ridge Associated Universities under contract with NASA. Part of this work was done at Jet Propulsion Laboratory, California Institute of Technology, under a contract with the National Aeronautics and Space Administration (80NM0018D0004). \textcopyright \ 2025. All rights reserved.

\bibliography{main} 
\bibliographystyle{aasjournal} 

\section{Appendix}
\label{sec:appendix}

\begin{figure*}[h]
    \centering
    \includegraphics[width=0.95\textwidth]{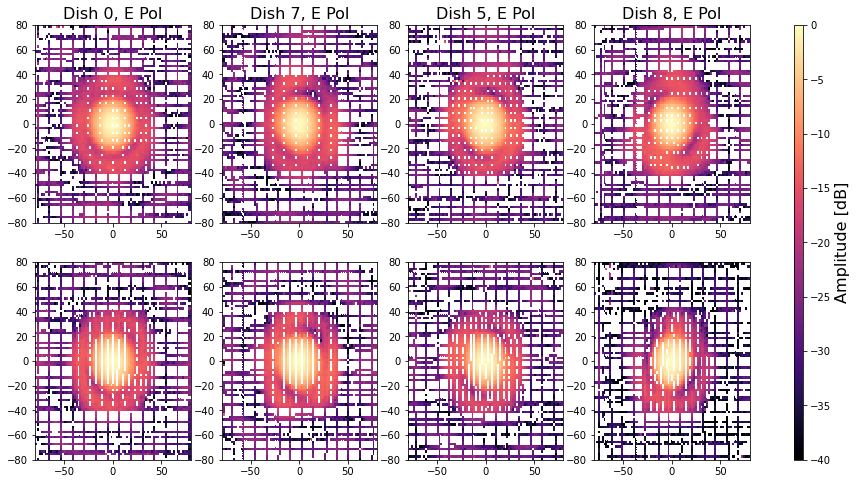}
    \caption{Panels show the beam maps from all inputs, at frequency 448\,MHz, in 2D Cartesian coordinates. In this gridding, an asymmetry in the dip feature between the main beam and first sidelobe is clearly present in all inputs. Dish 0, North-polarization appears to have a symmetric dip which is circular, while both inputs on Dish 7 and Dish 5 have an asymmetry that is aligned $\sim 35^{\circ}$ clockwise from North. Both inputs on Dish 8 have an asymmetry that is aligned $\sim 45^{\circ}$ counterclockwise from North. }
    \label{fig:xy_beam}
\end{figure*}

\begin{figure*}[b]
    \centering
    \includegraphics[width=0.95\textwidth]{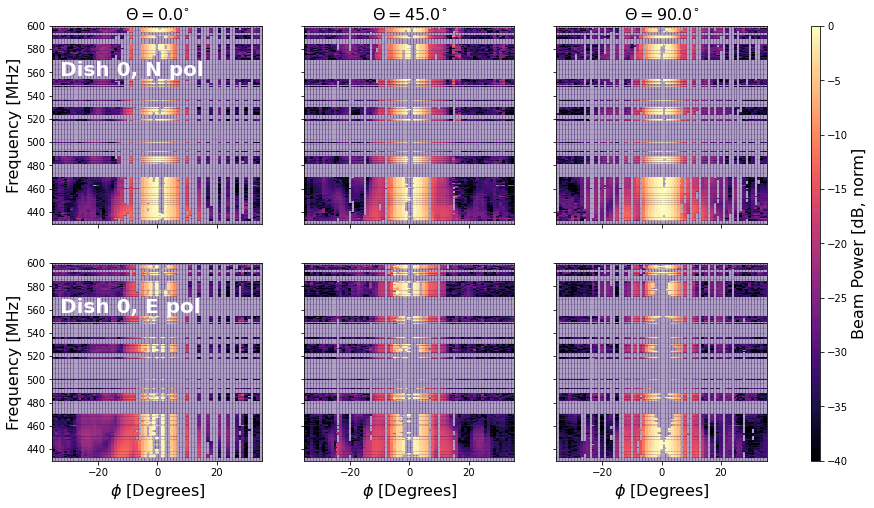}
    \caption[]{Each panel shows the evolution of the beam with frequency for three slices (left to right: $\theta=0^{\circ},45^{\circ},90^{\circ}$) for Dish 0, (top is North-polarized input, bottom is East-polarized input). Gaps are a result of the frequency selection described in Section~\ref{sec:processing}. The beam width of the main beam is slightly larger at lower frequencies, and the sidelobes shift outwards and widen with it. The amplitude of the first sidelobe is similar across frequencies, the amplitude of the second sidelobe is more variable in both angular location (this is also apparent in Figure~\ref{fig:theta_phi_beam}) and as a function of frequency.}
    \label{fig:freq_phi_beam}
\end{figure*}

\begin{figure*}[h]
    \centering
    \includegraphics[width=0.9\textwidth]{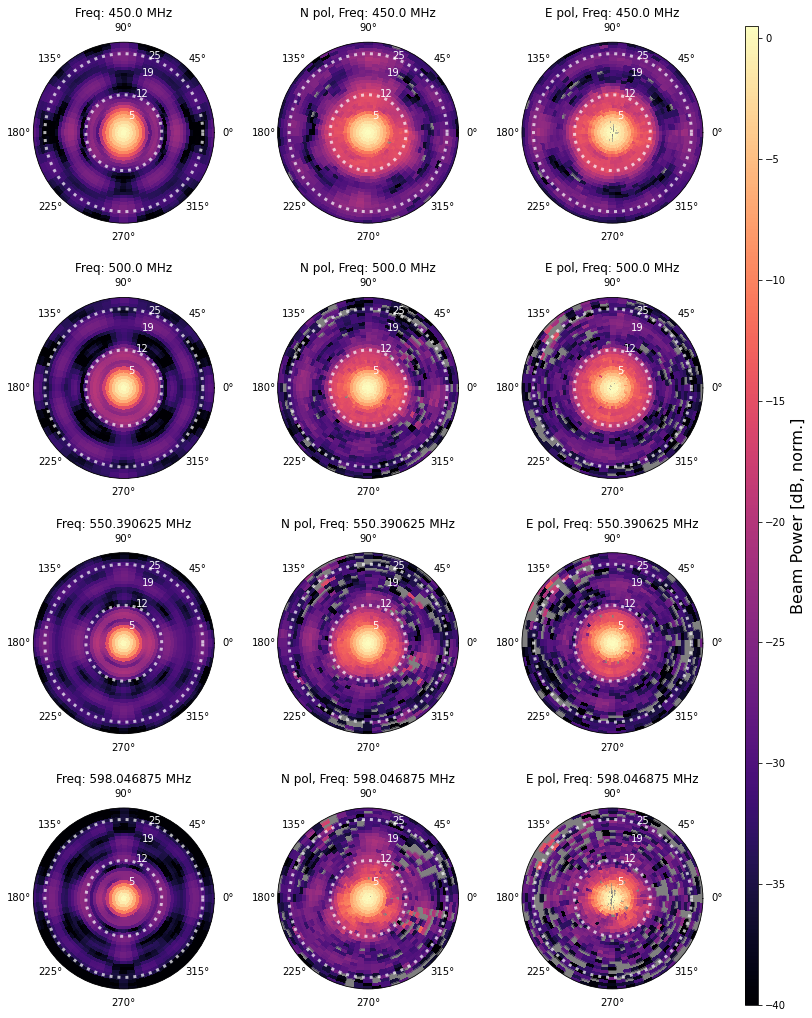}
    \caption[]{Simulated (left column) and measured (right columns) beams for the TONE dishes at four frequencies (from top to bottom: 450, 500,  550, and 600~MHz). The left column shows beams simulated using the CST Studio Suite software. The center and right columns show the measured beams for the polarizations nominally oriented North-South and East-West, respectively. Dashed lines have been drawn at the approximate locations of the first and second sidelobes in the measured data (12$^{\circ}$ and 25$^{\circ}$). }
    \label{fig:simsmaps}
\end{figure*}


\end{document}